\documentclass[fleqn,usenatbib]{mnras}

\usepackage{newtxtext,newtxmath}

\usepackage[T1]{fontenc}

\DeclareRobustCommand{\VAN}[3]{#2}
\let\VANthebibliography\thebibliography
\def\thebibliography{\DeclareRobustCommand{\VAN}[3]{##3}\VANthebibliography}

\usepackage{graphicx}	
\usepackage{amsmath}	
\usepackage{subfig}
\usepackage{CJK}
\usepackage{babel}
\usepackage{xcolor}

\title[WD\,0141$-$675: A case study]{WD\,0141$-$675: A case study on how to follow-up astrometric planet candidates around white dwarfs}

\author[L. K. Rogers et al.]{
Laura K. Rogers,$^{1}$\thanks{E-mail: laura.rogers@ast.cam.ac.uk} 
John Debes,$^{2}$
Richard J. Anslow,$^{1}$ 
Amy Bonsor,$^{1}$
S. L. Casewell,$^{3}$ 
\newauthor Leonardo~A.~Dos Santos,$^{2}$
Patrick Dufour,$^{4}$
Boris G\"{a}nsicke,$^{5}$
Nicola Gentile Fusillo,$^{6}$ 
Detlev Koester,$^{7}$ 
\newauthor Louise Dyregaard Nielsen,$^{8}$
Zephyr Penoyre,$^{9}$ 
Emily L. Rickman,$^{10}$ 
Johannes Sahlmann,$^{11}$ 
\newauthor Pier-Emmanuel Tremblay,$^{5}$
Andrew Vanderburg,$^{12}$
Siyi Xu \begin{CJK*}{UTF8}{gbsn}(许\CJKfamily{bsmi}偲\CJKfamily{gbsn}艺)\end{CJK*},$^{13}$ 
Erik Dennihy,$^{14}$
Jay Farihi,$^{15}$
\newauthor J. J. Hermes,$^{16}$
Simon Hodgkin,$^{1}$
Mukremin Kilic,$^{17}$
Piotr M. Kowalski,$^{18}$ 
Hannah Sanderson$^{19}$ and
\newauthor Silvia Toonen$^{20}$ 
\\
$^{1}$Institute of Astronomy, University of Cambridge, Madingley Road, Cambridge, CB3 0HA, UK \\
$^{2}$Space Telescope Science Institute, Baltimore, MD, USA\\
$^{3}$Centre for exoplanet research, School of Physics and Astronomy, University of Leicester, University Road, Leicester, LE1 7RH, UK \\
$^{4}$ D\'epartement de Physique, Universit\'e de Montr\'eal, C.P. 6128, Succ. Centre-Ville, Montr\'eal, Qu\'ebec H3C 3J7, Canada \\
$^{5}$ Department of Physics, University of Warwick, Coventry, CV4 7AL, UK \\
$^{6}$ European Space Agency, European Space Astronomy Centre (ESAC) \\
$^{7}$ Institut f$\ddot{u}$r Theoretische Physik und Astrophysik, Universit$\ddot{a}$t Kiel, 24098 Kiel, Germany  \\ 
$^{8}$ Karl-Schwarzschild-Stra{\ss}e 2, 85748 Garching bei M$\ddot{u}$nchen, Germany \\
$^{9}$ Leiden Observatory, Leiden University, PO Box 9513, 2300 RA, Leiden, The Netherlands \\ 
$^{10}$ European Space Agency (ESA), ESA Office, Space Telescope Science Institute, 3700 San Martin Drive, Baltimore, MD 21218, USA \\ 
$^{11}$ RHEA Group for the European Space Agency (ESA), European Space Astronomy Centre (ESAC), Camino Bajo del Castillo s/n, 28692 Villanueva de la Ca\~nada, Madrid, Spain \\ 
$^{12}$ Department of Physics and Kavli Institute for Astrophysics and Space Research, Massachusetts Institute of Technology, Cambridge, MA 02139, USA \\ 
$^{13}$ Gemini Observatory/NSF's NOIRLab, 670 N. A'ohoku Place, Hilo, HI 96720, USA \\ 
$^{14}$ Rubin Observatory Project Office, 950 N. Cherry Ave., Tucson, AZ 85719, USA \\
$^{15}$ Department of Physics and Astronomy, University College London, London, WC1E 6BT, UK \\
$^{16}$ Department of Astronomy \& Institute for Astrophysical Research, Boston University, 725 Commonwealth Avenue, Boston, MA 02215, USA \\
$^{17}$ Homer L. Dodge Department of Physics and Astronomy, University of Oklahoma, 440 W. Brooks St., Norman, OK 73019, USA \\ 
$^{18}$ Institute of Energy and Climate Research (IEK-13), Forschungszentrum J$\ddot{u}$lich, Wilhelm-Johnen-Stra{\ss}e, J$\ddot{u}$lich 52425, Germany \\ 
$^{19}$ Department of Earth Sciences, University of Oxford, South Parks Road, Oxford OX1 3AN, UK \\
$^{20}$ Anton Pannekoek Institute for Astronomy, University of Amsterdam, Science Park 904, 1098 XH Amsterdam, Netherlands \\
}

\date{Accepted 2023 October 6. Received YYY; in original form ZZZ}

\pubyear{2023}

\begin{document}
\label{firstpage}
\pagerange{\pageref{firstpage}--\pageref{lastpage}}
\maketitle

\begin{abstract}

This work combines spectroscopic and photometric data of the polluted white dwarf WD\,0141$-$675 which has a now retracted astrometric super-Jupiter candidate and investigates the most promising ways to confirm \textit{Gaia} astrometric planetary candidates and obtain follow-up data. Obtaining precise radial velocity measurement for white dwarfs is challenging due to their intrinsic faint magnitudes, lack of spectral absorption lines, and broad spectral features. However, dedicated radial velocity campaigns are capable of confirming close in giant exoplanets (a few M$_{\textrm{Jup}}$) around polluted white dwarfs, where additional metal lines aid radial velocity measurements. Infrared emission from these giant exoplanets is shown to be detectable with \textit{JWST} MIRI and will provide constraints on the formation of the planet. Using the initial \textit{Gaia} astrometric solution for WD\,0141$-$675 as a case study, if there were a planet with a 33.65\,d period or less with a nearly edge on orbit, 1) ground-based radial velocity monitoring limits the mass to $<$\,15.4\,M$_{\textrm{Jup}}$, and 2) space-based infrared photometry shows a lack of infrared excess and in a cloud-free planetary cooling scenario, a sub-stellar companion would have to be $<$\,16\,M$_{\textrm{Jup}}$ and be older than 3.7\,Gyr. These results demonstrate how radial velocities and infrared photometry can probe the mass of the objects producing some of the astrometric signals, and rule out parts of the brown dwarf and planet mass parameter space. Therefore, combining astrometric data with spectroscopic and photometric data is crucial to both confirm, and characterise astrometric planet candidates around white dwarfs.

\end{abstract}




\section{Introduction}


Exoplanets are known to be ubiquitous, with over 5000 discovered around main sequence stars; however, few planets have been confirmed around white dwarf stars. White dwarfs are the final products of stellar evolution for low- and intermediate- mass stars with an initial main sequence mass less than 9--12\,M$_{\odot}$ \citep[e.g.][]{Lauffer2018new,Althaus2021formation}; roughly 97 percent of the stars in the Milky Way will evolve to white dwarfs \citep{fontaine2001potential}. The accurate \textit{Gaia} parallax measurements for stars in the Milky Way has enabled the number of white dwarfs discovered to increase by an order of magnitude, with the most recent catalogue reporting 359,000 high-confidence white dwarf candidates \citep{gentile2021catalogue}. 

There is an ever-increasing number of remnant planetary systems discovered around white dwarf stars. The most common observations are metal pollution in the white dwarfs' photosphere which is attributed to the accretion of planetesimals that have been tidally disrupted when on star-grazing orbits. This is observed in 25--50 percent of white dwarfs cooler than 25,000\,K, where radiative levitation effects are negligible so the pollution source must be external \citep{zuckerman2003metal, zuckerman2010ancient, koester2014frequency, wilson2019unbiased}. Observations of the circumstellar environment reveal: infrared excesses from circumstellar dust in 1.5--4 percent of white dwarfs \citep[e.g.][]{wilson2019unbiased}, double peaked gaseous emission features in 0.067 percent of white dwarfs \citep{manser2020frequency}, and transiting debris around of order 10 white dwarfs \citep[e.g.][]{vanderburg2015disintegrating}. However, there are still very few confirmed planets around white dwarfs, and they are either inferred to be present, or remain unconfirmed \citep{sigurdsson2003young, gansicke2019accretion, blackman2021jovian}. The most notable candidate is WD\,1856+534b, a giant transiting planet candidate found from TESS data on an orbit of 1.4\,d \citep{vanderburg2020giant}. The tidal disruption scenario as the origin of the observed metal pollution requires the presence of surviving large planets in the outer system, capable of scattering smaller rocky bodies within the tidal disruption radius of the white dwarf \citep[e.g.][]{debes2002there}. In order for 25--50 percent of white dwarfs to be polluted, surviving giant planets must therefore be relatively common. However, planets around white dwarfs remain difficult to detect due to, for example, low transit probabilities or few/no narrow absorption lines for  precise radial velocity measurements \citep[e.g.][]{faedi2011detection,maxted2000radial,belardi2016decam}. 

There are more detections of white dwarfs with sub-stellar companions than there are detections of planets around white dwarfs, however, these are still rare with estimates of an occurrence rate of approximately 0.5 percent \citep{steele2011white}. This low number is partly as a result of sub-stellar companions not surviving the evolution of the host star due to their masses and orbital locations \citep{Nelemans1998formation,Nordhaus2010tides,Mustill2013main}. The commonly accepted origin of close white dwarf-sub stellar binaries is post common-envelope evolution, where one star in a binary system evolves to a red giant and overflows its Roche lobe forming a common envelope around the binary \citep[e.g.][]{lagos2021wd}. Drag causes orbital angular momentum to be lost, resulting in the binary separation decreasing. If the envelope is ejected, this leaves a white dwarf in a binary with a close companion \citep[e.g.][]{maxted2006survival}. It is unknown whether pollution is correlated or uncorrelated with the presence of a close sub-stellar companion. Only one white dwarf-brown dwarf system shows metal pollution; \citet{farihi2017circumbinary} reported a white dwarf-sub stellar binary, where a circumbinary dust disc orbiting a polluted white dwarf (SDSS\,J155720.77+091624.6) was identified with a brown dwarf companion on a 2.27 hour orbit. 

The exquisite astrometric capabilities of \textit{Gaia} opens a new window of opportunity to discover planets around white dwarfs; with the release of \textit{Gaia} DR5 around 10 gas giants on wide orbits (0.03$-$13\,M$_{\textrm{Jup}}$ with semi-major axes 1.6$-$3.91\,au) are expected to be discovered around white dwarfs \citep{perryman2014astrometric,sanderson2022can}. The first astrometric orbital solutions were released as part of the \textit{Gaia} DR3 \citep{vallenari2022gaia}. This presented the first astrometric planet candidate around a white dwarf \citep{collaboration2022gaia}, a super-Jupiter around \textit{Gaia} DR3 4698424845771339520 (WD\,0141$-$675). Due to its proximity to the Sun (9.72\,pc) and its brightness, WD\,0141$-$675 is a well studied white dwarf with archival photometric and spectroscopic data covering the ultraviolet through to mid-infrared wavelengths. It was first classified as a DA white dwarf in \citet{hintzen1979observations}, and high-resolution spectra later revealed the presence of calcium H and K lines, a sign that this white dwarf is polluted with heavy metals \citep{debes2010results}. Both accretion from the ISM and intermediate absorption from the ISM are unlikely to explain the pollution as WD\,0141$-$675 is 9.72\,pc away and lies within the local bubble. As it is nearby, bright, and polluted it was included in the search for astrometric companions with \textit{Gaia} DR3 \citep{holl2022gaia}. The astrometric data was better fitted by including a companion instead of a single star. The \textit{Gaia} DR3 non-single star orbital solution had a companion mass of 9.26\,M$_{\textrm{Jup}}$ and a period of 33.65\,d, see Table \ref{tab:WD-prop} for further details. In May 2023 it was announced that there were false positives in the \textit{Gaia} astrometric solutions, and  WD\,0141$-$675 no longer has a non-single star astrometric solution \footnote{\url{https://www.cosmos.esa.int/web/gaia/dr3-known-issues}}.

This work presents a case study using the nearby polluted white dwarf WD\,0141$-$675 to place constraints on the planetary material being accreted by the white dwarf and investigate how to follow-up astrometric candidate planets to confirm and characterise them. This paper presents a detailed photometric and spectroscopic analysis of WD\,0141$-$675 to probe what could be done to confirm and characterise astrometric planet candidates given an astrometric solution similar to that originally posed with a planetary mass of 9.26\,M$_{\textrm{Jup}}$ and a period of 33.65\,d. Section \ref{Spec_Obs} reports the optical and ultraviolet spectroscopic observations that were used to obtain the abundance of the material polluting the white dwarf, constrain the accretion rate onto the white dwarf, and put limits on the composition of the parent body accreted, as derived in Section \ref{Abundances}. Section \ref{SED-Modelling} reports the full spectral energy distribution (SED) with details on how to calculate limits to emission from planetary companions calculated using planetary cooling models. Radial velocities were calculated using the optical spectra to put constraints on the mass of the planet candidate, as described in Section \ref{RV}. Section \ref{Simulations} investigates the origin of the pollutant material in the photosphere of the white dwarf. Finally, the discussions and conclusions on how to use this work to follow up astrometric planet candidates are presented in Sections \ref{Discussion} and \ref{Conclusions}. 

\begin{table}
	\centering
	\caption{Properties of WD\,0141$-$675 from \textit{Gaia} DR3 and the astrometric orbital solution for the (now retracted) candidate planet.}
	\label{tab:WD-prop}
	\begin{tabular}{ll} 
		\hline
		\textbf{\textit{Gaia} DR3 Number:} & 4698424845771339520 \\
        \hline
        RA & 01:43:00.98  \\
        DEC & $-$67:18:30.35 \\
        D (pc) & 9.72 \\
        G band (mag) & 13.69 \\
        G$_{\textrm{BP}}$ band (mag) & 13.94 \\
        G$_{\textrm{RP}}$ band (mag) & 13.29 \\
        SpT & DAZ$^{a}$ \\
        RUWE & 1.049 \\
		\hline	
       \textbf{Subasavage et al. (2017)$^{b}$:} \\ \vspace{0.5mm}
        T${_{\textrm{eff}} }$ (K) & 6380 (120) \\ \vspace{0.5mm}
        $\log(g)$ (cm\,s$^{-2}$) & 7.97 (0.03) \\ \vspace{0.5mm}
        $\tau _{\textrm{cool}}$ (Gyr) & 1.94\,$^{+0.13}_{-0.12}$$^{d}$ \\ \vspace{0.5mm}
        Total age (Gyr) & 8.02\,$^{+3.62}_{-2.99}$$^{d}$ \\ \vspace{0.5mm}
        M (M$_{\odot}$) & 0.58 (0.02)$^{d}$ \\  \vspace{0.5mm}
        log(q)$^{c}$ & $-$7.59 \\ \\
        \textbf{Spectroscopic method (this work):} \\ \vspace{0.5mm}
        T${_{\textrm{eff}} }$ (K) & 6421 (90) \\  \vspace{0.5mm}
        $\log(g)$ (cm\,s$^{-2}$) & 8.10 (0.04) \\ \vspace{0.5mm}
        $\tau _{\textrm{cool}}$ (Gyr) & 2.30\,$^{+0.27}_{-0.18}$$^{d}$ \\ \vspace{0.5mm}
        Total age (Gyr) & 4.03\,$^{+2.38}_{-0.59}$$^{d}$ \\ \vspace{0.5mm}
        M (M$_{\odot}$) & 0.65 (0.03)$^{d}$ \\ \vspace{0.5mm}
        log(q)$^{c}$ & $-$7.80 \\ \\
        \textbf{Photometric method (this work):} \\ \vspace{0.5mm}
        T${_{\textrm{eff}} }$ (K) & 6321 (50) \\ \vspace{0.5mm}
        $\log(g)$ (cm\,s$^{-2}$) & 7.97 (0.03) \\ \vspace{0.5mm}
        $\tau _{\textrm{cool}}$ (Gyr) & 1.98\,$^{+0.10}_{-0.09}$$^{d}$ \\ \vspace{0.5mm}
        Total age (Gyr) & 8.06\,$^{+3.64}_{-2.98}$$^{d}$ \\ \vspace{0.5mm}
        M (M$_{\odot}$) & 0.57 (0.02)$^{d}$ \\    \vspace{0.5mm}  
        log(q)$^{c}$ & $-$7.55 \\
        \hline
        Planet Period (d) & 33.65\,$\pm$\,0.05$^{e}$ \\
        Planet Mass (M$_\textrm{Jup}$) & 9.26$_{-1.15}^{+2.64}$$^{e}$ \\
        Inclination (deg) & 87.0\,$\pm$\,4.1$^{e}$ \\
        \hline
	\end{tabular}
    \vspace{-2mm}
 	\small \begin{flushleft}
      \item \textbf{Notes:} 
      \item $^{a}$SpT from \citet{debes2010results}.
      \item $^{b}$White dwarf parameters from \citet{subasavage2017solar} derived using photometry.
      \item $^{c}$ $\textrm{q}=log_{\textrm{10}}(M_{\textrm{CVZ}}/M_{\textrm{WD}})$, the mass ratio of the convective zone of the white dwarf, $M_{\textrm{CVZ}}$, to the total mass of the white dwarf, $M_{\textrm{WD}}$.
      \item $^{d}$ Total age, cooling age ($\tau _{\textrm{cool}}$), and mass are calculated using \textsc{wdwarfdate} \citep{Kiman2022wdwarfdate}.
      \item $^{e}$Planet parameters from \citet{collaboration2022gaia}, inferred from Monte-Carlo resampling, the median of the Planet Mass is 9.26\,M$_\textrm{Jup}$, and the mode is 8.3\,M$_\textrm{Jup}$. The posterior distribution on the mass overlaps into the brown dwarf mass regime. 
       \end{flushleft}
\end{table}

\section{Spectroscopic Observations} \label{Spec_Obs}

\subsection{MIKE/Magellan}
WD\,0141$-$675 was observed five times with the Magellan Inamori Kyocera Echelle (MIKE) spectrograph \citep{bernstein2003mike} on the 6.5\,m Magellan Clay Telescope at Las Campanas Observatory between 2008 -- 2010  (PI: Debes). The MIKE spectrograph is a double echelle spectrograph with two arms covering wavelengths of 3350--5000\,\AA\ (blue) and 4900--9500\,\AA\ (red). The data were obtained under varying atmospheric conditions with the 0.7\,$\times$\,5" slit. This gave a spectral resolution of $\sim$\,35,000 for the red arm and 46,000 for the blue arm. The spectra were binned at 2\,$\times$\,2 and had a slow readout speed. The data were reduced with the standard Carnegie Python MIKE pipeline included extraction, flat-fielding and wavelength calibration \citep{kelson2000evolution, kelson2003optimal}. Due to the conditions, this gave a signal-to-noise ratio (S/N) per pixel of 8--43, as reported in Table \ref{tab:RVs}. To determine the pollutant abundance, these data were stacked together, weighted by their S/N, to achieve a final stacked spectrum with a S/N of 50--60 at the continuum around the calcium K line (3933\,\AA). These data were corrected for radial velocity shifts before they were stacked.

\subsection{VLT X-shooter}

WD\,0141$-$675 was observed with the echelle spectrograph X-shooter \citep{vernet2011x} on Unit Telescope 3 of the Very Large Telescope (VLT) at Paranal Observatory, Chile. X-shooter enables simultaneous observations in the 3 arms: UVB (3000--5595\,\AA), VIS (5595--10240\,\AA) and NIR (10240--24800\,\AA). Observations of WD\,0141$-$675 have been taken in two modes: STARE, where a spectrum is taken using a fixed point on the sky, and NOD mode, where the telescope is nodded between two positions to reduce the affect of the sky background, important for NIR observations. The NOD mode observations were taken on 2017-07-27 with run ID 099.D-0661 (PI: G\"{a}nsicke) with a nodding throw length of 5 arcsecs, and the STARE mode observations were taken on 2019-12-05 with run ID 1103.D-0763 (PI: G\"{a}nsicke). For all observations a 1.0, 0.9 and 0.9 arcsec slit width are used for the UVB, VIS and NIR arms, this gives an optimal resolution ($\lambda / \Delta \lambda $) of 5400, 8900, and 5600 respectively. The data reduction was performed using the standard procedures within the \textsc{Reflex} reduction tool\footnote{\url{http://www.eso.org/sci/software/esoreflex/}} developed by ESO with minor changes adopted to optimise the spectral extraction \citep{Freudling2013XShReduction}. Telluric line removal was performed on the reduced spectra using \textsc{Molecfit} \citep{smette2015molecfit,kausch2015molecfit}. This gave a S/N per pixel of 170 and 120, as reported in Table \ref{tab:RVs}.

\subsection{\textit{Hubble} STIS }

WD\,0141$-$675 was observed with the Space Telescope Imaging Spectrograph (STIS) on the \textit{Hubble Space Telescope} on 2016 May 04 (PI: G\"{a}nsicke). The G230L grating was used with a central wavelength of 2376\,\AA, this gives wavelength coverage of 1570--3180\,\AA\, and a resolving power of 500. The exposure time was 2293\,s giving a S/N of 70.

\section{SED and Spectral Modelling} \label{SED-Modelling}

\subsection{White dwarf parameters} \label{WD-params}
\begin{figure}
	\includegraphics[width=\columnwidth]{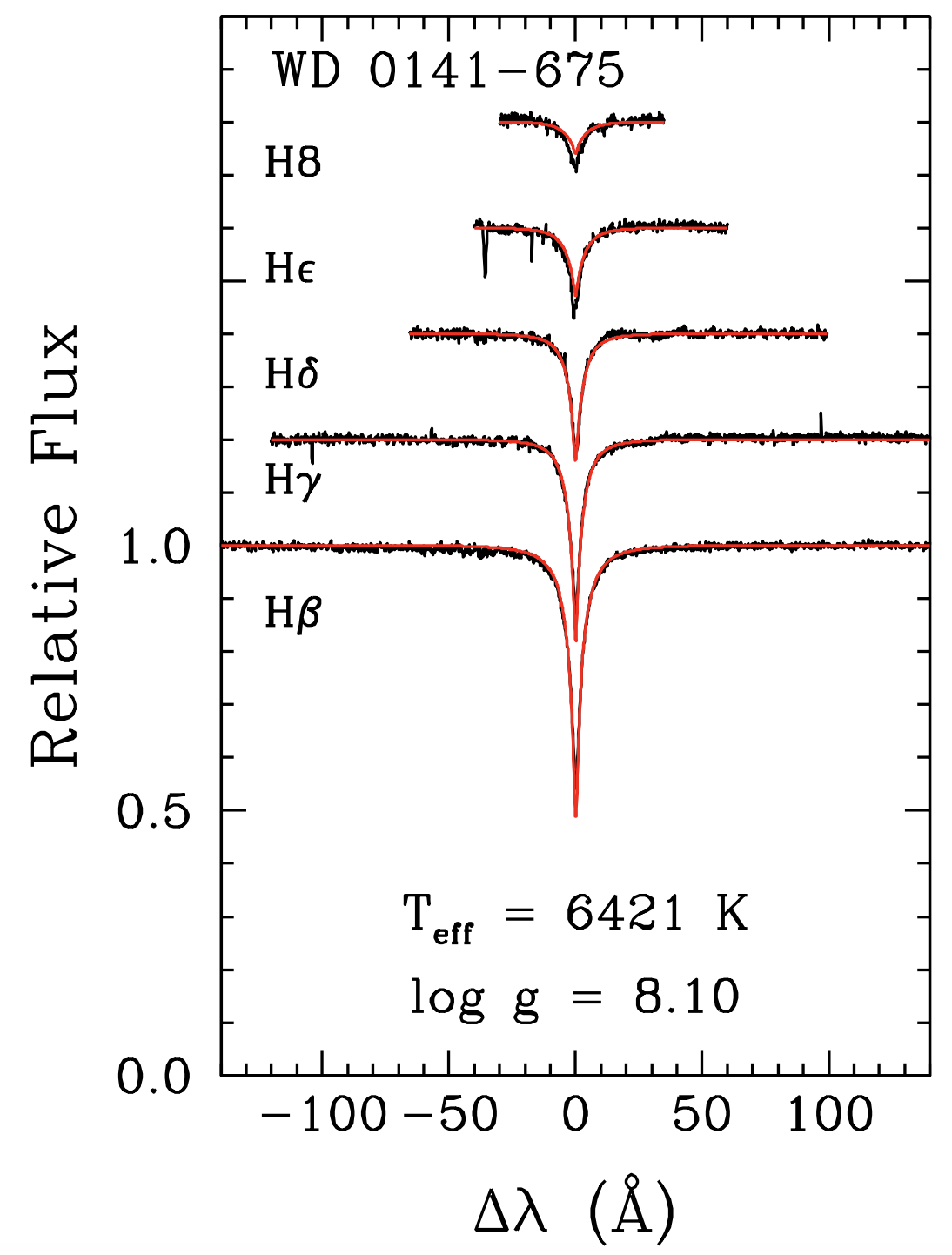}
    \caption{Normalised sections of the X-shooter spectrum for WD\,0141$-$675 showing the Balmer lines vertically offset from one another. The best fitting white dwarf model from the spectroscopic method (see Section \ref{WD-params} for further details) is over-plotted in red with an effective temperature of 6421\,K and $\log(g)$ of 8.10.}
    \label{fig:WD-fits}
\end{figure}

Two methods were used to derive the white dwarf parameters: the first fitted white dwarf models to the pressure broadened hydrogen spectral lines (the spectroscopic method), and the second fitted white dwarf models to broad band photometry (the photometric method). A description of these techniques, the models, and minimisation methods used are discussed in detail in \citet{Bergeron2019measurement, genest2019comprehensive}. The spectroscopic fit to the hydrogen absorption lines is shown in Fig.\,\ref{fig:WD-fits}, with an effective temperature of 6421\,$\pm$\,90\,K and $\log(g)$ of 8.10\,$\pm$0.04. The adopted uncertainties are from \citet{Liebert2005formation}, giving 1.4 percent in effective temperature and 0.04 dex in $\log(g)$; this does not take into account systematic errors from the model atmospheres. For the photometric method, \textit{GALEX} NUV, SkyMapper $uvgriz$ \citep{wolf2018skymapper} and 2MASS $JHK$ photometry \citep{cutri2003vizier} were fitted. Parallaxes from \textit{Gaia} were required to obtain $\log(g)$ and evolutionary models described in \citet{Bedard2020spectral} were used to derive the mass of the white dwarf. Reddening is not considered as it only becomes important for objects $>$ 100\,pc. The best fitting photometric effective temperature and $\log(g)$ were 6321\,$\pm$\,50\,K and 7.97\,$\pm$\,0.03 respectively (Table \ref{tab:WD-prop}), with 1\,$\sigma$ errors calculated separately using a reduced $\chi ^{2}$ analysis. The derived values for the spectroscopic and photometric methods are reported in Table \ref{tab:WD-prop}. Previously derived photometric parameters from \citet{subasavage2017solar} are reported for reference; it should be noted that these parameters were derived prior to the \textit{Gaia} parallax measurements. The total ages derived from the photometric and spectroscopic parameters are different by a factor of two, this has a significant impact on the detectability and properties of planets around these stars. 

\subsection{Spectral Energy Distribution}\label{sect:SED_model}

In order to assess the Spectral Energy Distribution (SED) of WD\,0141$-$675, photometry was collated from \textit{GALEX} in the NUV band, SkyMapper in the optical ($uvgriz$), 2MASS for the near-infrared photometry $JHK$ bands, and \textit{Spitzer} for the mid infrared \citet{kilic2008first}. The \citet{kilic2008first} data were used for all \textit{Spitzer} bands except for the 4.5\,$\mu$m band, where data from programme 60161 (PI: Burleigh) was used due to the higher S/N. The IRAC channel 2 (4.5\,$\mu$m) had 20 frames taken, each with an exposure time of 30\,s. These data were reduced using the \textsc{Mopex} pipeline, and aperture photometry performed using Apex and using the recommended pixel to pixel and array location corrections for the IRAC warm mission, resulting in a S/N of 210. 

\begin{figure*}
	\includegraphics[width=0.9\textwidth]{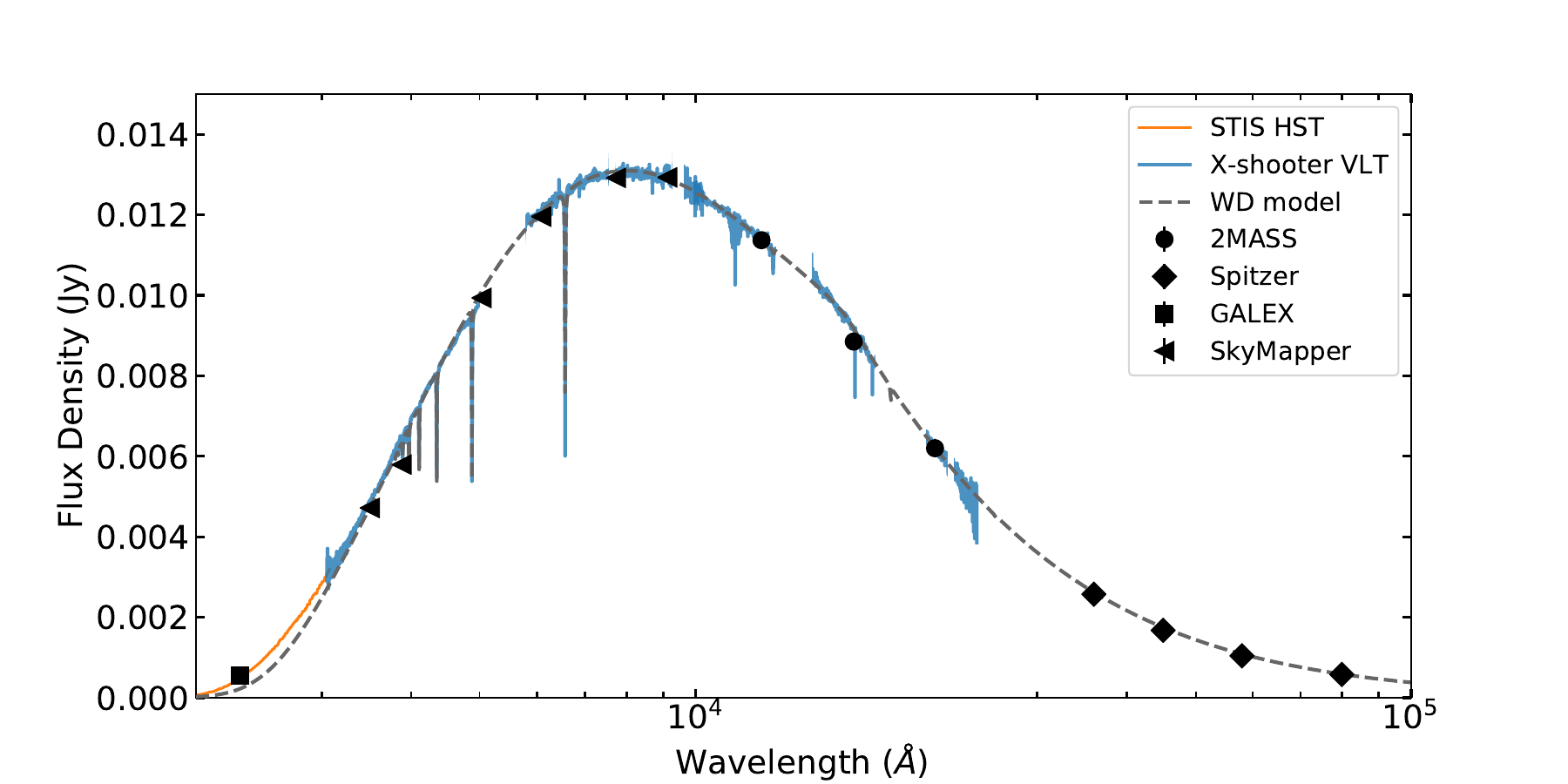}
    \caption{SED of WD\,0141$-$675. The spectra are from \textit{HST} STIS (orange), and VLT X-shooter (blue) which has been smoothed for clarity. The ultraviolet photometry is from the \textit{GALEX} NUV band, the optical photometry is from SkyMapper, the near-infrared photometry ($JHK$) is from 2MASS, and the mid-infrared photometry from \textit{Spitzer} is taken from \citet{kilic2008first}. The white dwarf model assumes the parameters derived in Section \ref{SED-Modelling} using the photometric method, with a temperature of 6321\,K and $\log(g)$ of 7.97, and is overlaid as a grey dashed line. The spectra are scaled to this model to account for slit losses.}
    \label{fig:SED}
\end{figure*}

The full SED including the spectra from VLT X-shooter and \textit{HST} STIS, and the photometry as outlined above is shown in Fig.\,\ref{fig:SED}. The white dwarf model over-plotted on the data assumes the photometric temperature and $\log(g)$. The STIS and \textit{GALEX} data in the NUV lie systematically above the white dwarf models. Different white dwarf models were tested in an attempt to reproduce the observed ultraviolet-excess including: models at higher effective temperature, atmospheric models which included He and metal pollutants \citep{dufour2012detailed}, models which included 3D effects \citep{Tremblay2013spectroscopic}, and independent models that previous work used to successfully reproduce NUV through IRAC spectra of cool (T$_{\textrm{eff}}$\,$<$\,5600\,K) white dwarfs \citep{Saumon2014nuv}. None of the models reproduce the ultraviolet-excess. This may point to some minor problems with the existing models, however, this is not important for the analysis performed in this paper.

\subsection{Photometric limits to the candidate companion}

Assuming the \textit{Gaia} astrometric solution for WD\,0141$-$675 with the planet candidate having a period of 33.65\,d, the companion is an unresolved source that could cause an infrared excess \citep[e.g.][]{steele2011white}, if bright enough. Previous studies which assessed the SED of WD\,0141$-$675 reported no excess infrared emission  \citep{mullally2007spitzer,kilic2006mystery}. This work corroborates these claims with neither the \textit{Spitzer} photometry, nor the near-infrared X-shooter arm displaying excess infrared emission above what is expected from an isolated white dwarf. Therefore, strong constraints can be put on the emission from a candidate cool companion.

Planetary cooling models are used to place photometric limits on a sub-stellar companion with the properties identified by the \textit{Gaia} astrometric solution. It is difficult to precisely predict the spectra of planets around white dwarfs as the age is unknown, the upper atmosphere may have been altered during post-main-sequence evolution, or the companion could have gained significant mass during common envelope evolution with the host star. When assessing emission from planets around white dwarfs it is therefore crucial to consider a range of potential ages. The following considers two bounding cases in order to help place constraints on what might be possible to detect with existing observatories and to check for upper limits to existing mid-infrared photometry of this object from \textit{Spitzer} and \textit{WISE} \citep{mullally2007spitzer,kilic2006mystery,kilic2008first}. The first bounding case is that the object was rejuvenated by the post-main sequence \citep[e.g.][]{spiegel2012jupiter} (or formed during the start of the post main sequence), and therefore for this case study of WD\,0141$-$675 it would have a planetary cooling age of 2\,Gyr. This corresponds to the average cooling age of the white dwarf from the three sets of white dwarf parameters in Table\,\ref{tab:WD-prop} calculated from \textsc{wdwarfdate} \citep{Kiman2022wdwarfdate} using cooling models from \citet{Bedard2020spectral}. The second bounding case is that it formed in the protoplanetary disc and has been untouched by the stellar evolution of its host and it consequently has a cooling age equivalent to the total age of the system (equal to the white dwarf cooling age plus the time on the main sequence). The total age was calculated with \textsc{wdwarfdate} using the \citet{cummings2018white} initial-final mass relation. Given the uncertainties on total age from different white dwarf parameters and that the initial-final mass relation is less reliable for the mass range of WD\,0141$-$675 \citep{Heintz2022Testing}, a conservative bounding case of 10\,Gyr is used. To calculate the predicted planetary emission, \textsc{Sonora-Bobcat} cloud-free, sub-stellar atmosphere models were used \citep{marley2021sonora}. In the following analysis it is assumed that the object has both a solar metallicity and C/O ratio.

\begin{figure*}
	\includegraphics[width=0.8\textwidth]{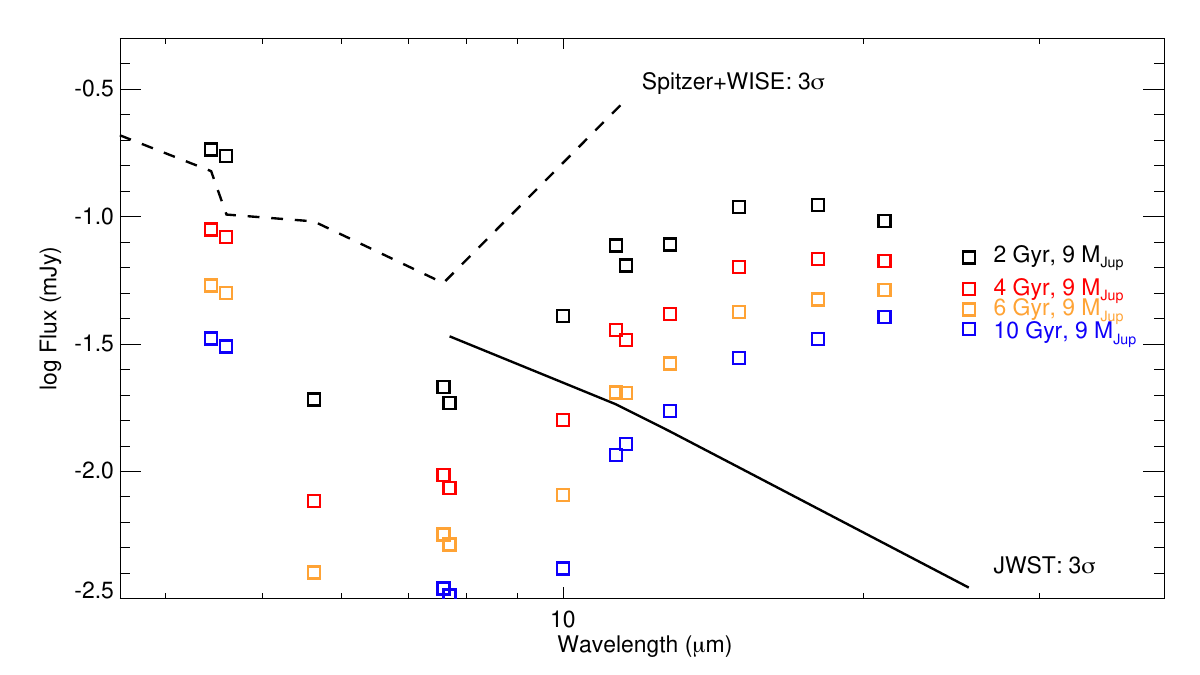}
    \caption{The expected emission from a 9.26\,M$_{\textrm{Jup}}$ planetary companion using \textsc{Sonora-Bobcat} models for planetary cooling ages between 2--10~Gyr as a function of wavelength for WD\,0141$-$675. The dashed black line shows the 3\,$\sigma$ upper limits from \textit{Spitzer}/\textit{WISE} photometry of WD\,0141$-$675; any models that lie above this should have been detected and can therefore be ruled out. From these limits, a companion temperature down to 330\,K is ruled out, equivalent to planetary cooling ages of up to 3.7\,Gyr. The predicted JWST 3\,$\sigma$ infrared excess limits (6 percent) are additionally shown as a black line; above 12\,$\mu$m \textit{JWST} MIRI can detect infrared emission from a 9.26\,M$_{\textrm{Jup}}$ cloud-free planet, regardless of whether it was 2 or 10~Gyr old. }
    \label{fig:Sonora}
\end{figure*}

The optimistic case is first considered where the companion accreted a significant amount of material during post-main sequence evolution such that it has a young cooling age. For this case study, at the best fit mass of 9.26\,M$_{\textrm{Jup}}$, a 2\,Gyr cooling age implies a T$_{\textrm{eff}}$=370\,K and $\log(g)$=4.3. For these objects, the companion is brightest in the mid-infrared at \textit{Spitzer} IRAC2 and \textit{WISE} W2, with an expected flux of 0.2~mJy. As seen in Fig.\,\ref{fig:Sonora}, there is high quality photometry of \textit{Spitzer} IRAC2 and \textit{WISE} W2 bands for WD\,0141$-$675 and these observations would have detected excesses above the photosphere at 3\,$\sigma$ significance for fluxes of 0.15 and 0.11~mJy respectively. Therefore, these limits are below the expected flux for a young companion with a mass of 9.26\,M$_{\textrm{Jup}}$, so a 2\,Gyr old planet with this mass is ruled out. The \textit{Spitzer} and \textit{WISE} upper limit corresponds to an upper mass limit of 7~M$_{\textrm{Jup}}$ for a 2~Gyr planet cooling age assuming a T$_{\textrm{eff}}$=330~K and a $\log(g)$=4.3.

Next, the case where the planet was unaffected by post main sequence evolution and matches the total age of the host white dwarf is considered. In this case for WD\,0141$-$675, a 10\,Gyr cooling age for a 9.26\,M$_{\textrm{Jup}}$ companion implies a T$_{\textrm{eff}}=250$~K and $\log(g)$=4.4. Again, the companion is brightest at IRAC2 and W2, with predicted fluxes of 0.026 and 0.024~mJy respectively, well below the upper limits and not detectable at these wavebands as seen in Fig.\,\ref{fig:Sonora}. The upper limits from \textit{Spitzer} and \textit{WISE} correspond to a 16\,M$_{\textrm{Jup}}$ companion for a 10\,Gyr planet age, assuming T$_{\textrm{eff}}$=330\,K and $\log(g)$=4.5. Therefore, if the planet has a clear atmosphere, this rules out companions more massive than 16~M$_{\textrm{Jup}}$. However, if the planet were cloudy, which would suppress the infrared flux, then no mass limits can be placed. 

Assuming the case study where WD\,0141$-$675 has a 9.26\,M$_{\textrm{Jup}}$ planetary mass companion, the \textit{Spitzer} and \textit{WISE} 3\,$\sigma$ upper limits corresponds to an age of 3.7\,Gyr, implying the planet candidate must be at least 3.7\,Gyr old. Figure\,\ref{fig:Sonora} demonstrates that \textit{JWST} is capable of detecting infrared emission from a 9.26\,M$_{\textrm{Jup}}$ planet of any age around WD\,0141$-$675. The 21\,$\mu$m \textit{JWST} 3\,$\sigma$ limit is equivalent to a planet just 16 percent brighter than Jupiter, when scaled to the distance of WD\,0141$-$675 (9.72\,pc). Therefore, as shown by this case study, \textit{JWST} is sensitive enough to detect infrared emission from giant planets around white dwarf stars.

\subsection{Photometric variability limits from existing surveys}

Photometric surveys conducted in optical and infrared wavebands can produce key information on the orbit of candidate companions. Optical photometric surveys can put constraints on transiting planets, and if detected can place stringent limits on planetary radius and period. This can enable follow-up spectroscopic observations of the transit or eclipse of the companion to constrain the structure and composition of the atmosphere. Infrared surveys may provide information on the emission from the planet \citep{Limbach2022MNRAS}, and time series data may detect variability in the planetary emission. This can be used to constrain the albedo of the companion.

For the case study of WD\,0141$-$675, assuming the \textit{Gaia} astrometric orbital solution and the errors associated with this, the candidate companion would have a 2.7 percent transit probability. The Transiting Exoplanet Survey Satellite (TESS, \citealt{ricker2015transiting}) obtained data of WD\,0141$-$675 in Sectors 1 and 2 (with a cadence of 2\,min); and Sector 29 (with a cadence of 20\,s), with each sector lasting approximately 27\,d. \textsc{Lightkurve} \citep{2018ascl.soft12013L} was used to assess whether any optical variability was observed that could be due to a transiting companion, or transiting tidally disrupted material. Lomb-Scargle periodograms were constructed which found no coherent, periodic variability (on any timescale ranging from minutes -- days) to a limit of at least 0.02 percent amplitude, and as each sector spans most of the planet period (each sector is 27\,d and the companion period was 33.65\,$\pm$\,0.05\,d). There is a 99.3 percent likelihood that a transit should have been observed with TESS, this likelihood takes into account the gaps in the middle of the sectors. Including ASAS3\footnote{\url{http://www.astrouw.edu.pl/asas/?page=asas3}}, ASASSN\footnote{\url{https://www.astronomy.ohio-state.edu/asassn/}}, and ATLAS \citep{2018PASPATLAS} ground-based data, this increases to 99.5 percent. If the candidate planet were transiting the host star, an inclination of at least 89.8 degrees would be required (R=1\,R$_{\textrm{Jup}}$). Therefore, with 99.5 percent confidence, an inclination of 89.8--90.2 degrees can be ruled out for the candidate companion. Therefore, photometric surveys can rule out a range of inclinations that would otherwise be compatible with the \textit{Gaia} astrometric solution.

Wide-field Infrared Survey Explorer (\textit{WISE}) and \textit{NEOWISE} observations from 2010--2020 were investigated using the unTimely catalogue from \citet{meisner2022untimely} to investigate whether the mid-infrared fluxes varied over this baseline. There are 16 detections of WD\,0141$-$675 in each of the 3.4 and 4.6\,$\mu$m bands taken over 4000\,d. The resulting mid-infrared lightcurves were found to be consistent with a straight line within the errors, so over 4000\,d there were no changes detected in the mid-infrared flux for WD\,0141$-$675.

\section{Radial Velocities} \label{RV}

For main sequence stars, radial velocities are crucial to detect exoplanets and determine their masses \citep[e.g.][]{mayor1995jupiter}. This is more difficult for white dwarfs as some white dwarfs have no spectral features (DCs) and if they do, the features are pressure broadened, making it difficult to obtain precise radial velocities. However, the astrometric candidates that \textit{Gaia} will identify around white dwarfs are in the gas giant regime (massive planets with large radial velocity amplitudes), and therefore, for some candidates they may show detectable radial velocity signatures. Additionally, if these planets are discovered around \textit{polluted} white dwarfs, the metal lines tend to be narrower enabling increased radial velocity precision. 

\begin{figure*}
	\includegraphics[width=0.8\textwidth]{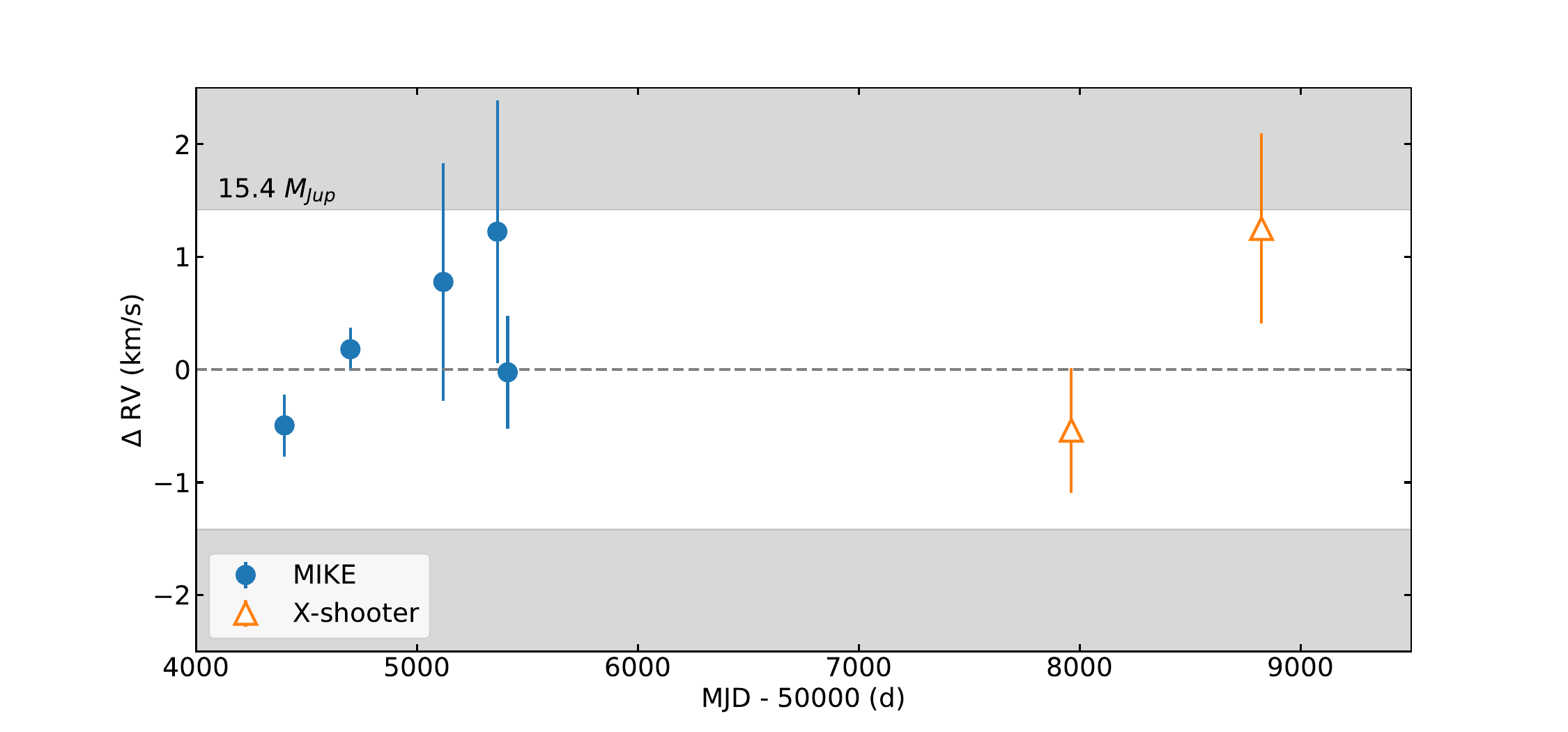}
    \caption{Seven radial velocity measurements of WD~0141$-$675 taken between 2008--2020 with MIKE (blue circle data points) and X-Shooter (orange triangle data points). For the MIKE and X-Shooter radial velocities separately, the weighted mean has been subtracted from each. Assuming a companion on a close to edge-on orbit, 87\,deg, with a period of 33.65\,d and that the five MIKE data points are sampling the full period, companions more massive than 15.4\,M$_{\textrm{Jup}}$ are ruled out for this case study and the grey regions represent radial velocity signatures that would result in a companion mass above this mass limit.}
    \label{fig:RVs}
\end{figure*}

WD~0141$-$675 is both bright and polluted with narrow metal lines, therefore, it is an ideal target for precision radial velocities. Radial velocities were calculated for the available mid and high-resolution optical spectra of WD~0141$-$675 as described in Section \ref{Spec_Obs}. The radial velocities were calculated by fitting a Gaussian profile to the core of the narrowest lines: the H\,$\alpha$ Balmer line, and the Calcium H and K metals lines from the polluted material. For each spectrum these lines were first measured to ensure they were detected at the 3\,$\sigma$ level before it was used for the radial velocity calculation. A Gaussian profile was fitted to the core of these lines using the \textsc{Idl} programme \textsc{Gaussfit} which gave the position of the center of the line and the estimated uncertainty on the fit parameters. From this, the weighted average of the line centers was found using the estimated uncertainties as the weights, and these were converted to a radial velocity per spectrum and are reported in Table \ref{tab:RVs}.  Figure\,\ref{fig:RVs} shows five epochs of radial velocities derived from photospheric calcium lines and hydrogen lines taken with the MIKE spectrograph between 2008--2010 and two epochs with the X-Shooter spectrograph in 2017 and 2020. The uncertainties are dominated by the stability of the instruments and for MIKE this is estimated by observations of another nearby white dwarf (1\,$\sigma = $ 0.50\,km\,s$^{-1}$). For X-shooter, \citet{parsons2017testing} report that the velocity precision is approximately 1\,km\,s$^{-1}$; measurements of sky emission lines in the VIS arm for WD\,0141$-$675 corroborates this precision. The rms of the radial velocities including the X-shooter and MIKE data for WD~0141$-$675 is 0.94\,km\,s$^{-1}$, and excluding the X-shooter data is 0.47\,km\,s$^{-1}$. Therefore, using the MIKE data the rms of the radial velocity data is 0.47\,$\pm$\,0.50\,km\,s$^{-1}$ and there is no evidence that the MIKE radial velocities have an additional radial velocity component due to a planetary companion. Taking this rms value as 1\,$\sigma$, the 3\,$\sigma$ upper limit for a companion is 15.4\,M$_{\textrm{Jup}}$; this upper limit assumes the \textit{Gaia} orbital period (33.65\,$\pm$\,0.05\,d), \textit{Gaia} inclination (87.0\,$\pm$\,4.1\,deg), and that these data are sampling the full period. Therefore, at the 3\,$\sigma$ limit, companions more massive than 15.4\,M$_{\textrm{Jup}}$ could be ruled out for this case study, shown as the shaded grey region in Fig.\,\ref{fig:RVs}.

\begin{figure*}%
    \centering
    \subfloat[\centering $i = 87.0\,\pm\,4.1$\,deg]{{\includegraphics[width=\columnwidth]{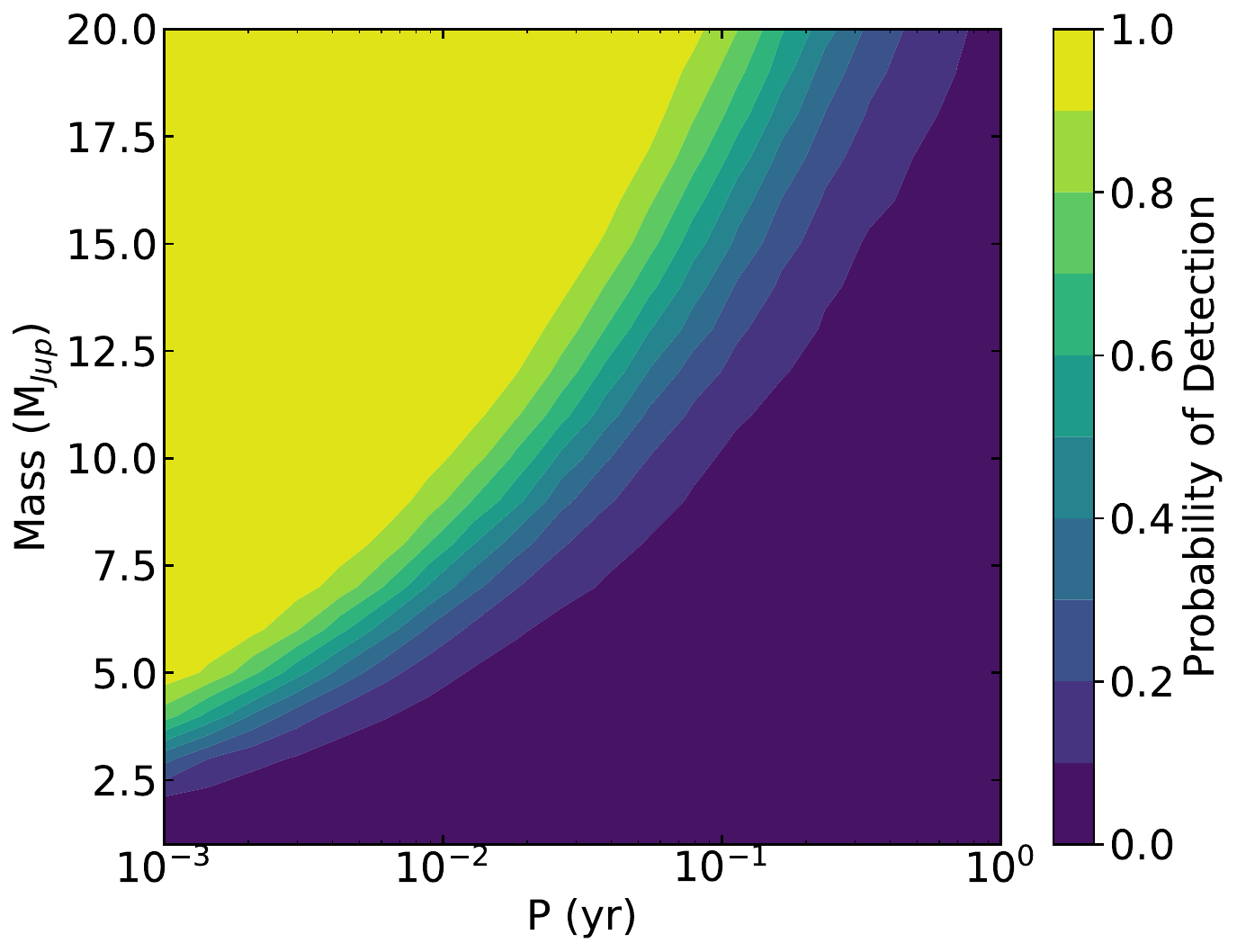} }}%
    \qquad
    \subfloat[\centering  $0 < i < 90$\,deg]{{\includegraphics[width=\columnwidth]{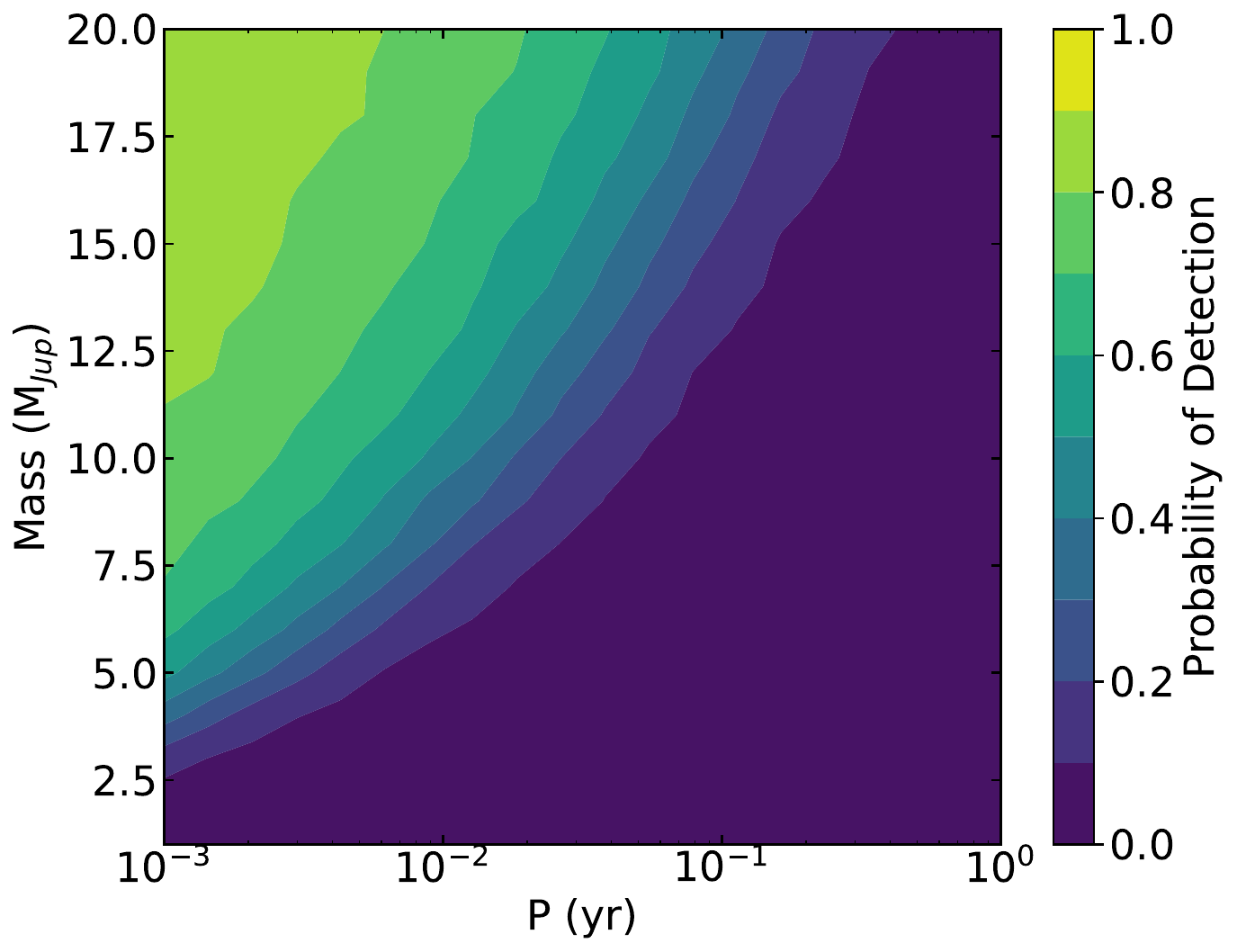} }}%
    \caption{Contour plots showing Monte Carlo models that calculate the probability of detecting a planet around WD\,0141$-$675 of a particular mass on an orbit with period, \textit{P}, assuming that five spectra were taken that sample the period with the radial velocity precision taken as the median of that of the MIKE datasets (0.5\,km s$^{-1}$). (a) Model assuming the inclination from the astrometric orbit $i = 87.0\,\pm\,4.1$\,deg, and (b) model assuming no prior information is known about the inclination. See Section \ref{RV} for further details.}%
    \label{fig:RVs-model}%
\end{figure*}

Monte Carlo models were used to show the range of parameter space of planets that can be ruled out based on a radial velocity campaign using MIKE data that sampled the period five times. The aim was to calculate the probability of detecting a planet with a particular mass and period on a circular orbit around a white dwarf of mass 0.6\,$M_{\odot}$ using the MIKE observations as the basis for the observing strategy. Two models were conducted, as shown in Fig.\,\ref{fig:RVs-model}, the first assuming an inclination of $i = 87.0\,\pm\,4.1$\,deg, and the second assuming no prior information on the inclination. For each mass and period, 5000 simulated observations were constructed, the first model drew random inclinations from a Gaussian distribution centered on 87.0 deg with a standard deviation of 4.1 deg, and the second drew random orbital inclination from a uniform distribution between 0 and 90 deg. This produced a radial velocity curve that was sampled five times, with each radial velocity measurement being drawn from a Gaussian centered on that velocity and a standard deviation given by the median radial velocity error from MIKE (0.5\,km\,s$^{-1}$). This data was then fitted with a radial velocity curve and if the signal could be detected at $>$ 3\,$\sigma$, where $\sigma$ is the spread in radial velocities expected for WD\,0141$-$675 (0.47\,km\,s$^{-1}$), then it is characterised as a planet detection. Figure\,\ref{fig:RVs-model} shows the results of the Monte Carlo models. If the orbital inclination were close to edge on, companions of mass $>$\,15\,M$_{\textrm{Jup}}$ with periods of $<$\,15\,d and $>$\,6.5\,M$_{\textrm{Jup}}$ with periods of $<$\,1\,d can be ruled out confidently. However, if the astrometric signal resulted in an unconstrained inclination then none of the parameter space can be ruled out with certainty.

\begin{table}
	\centering
	\caption{Radial velocities (RV) from MIKE and X-shooter. S/N measured per pixel at the continuum around the calcium K line (3933\AA). }
	\label{tab:RVs}
	\begin{tabular}{lllll} 
		\hline
		Instrument & MJD (d) & S/N  & RV (km\,s$^{-1}$) \\
        \hline
        MIKE & 54698.691 & 43.0 & 63.19\,$\pm$\,0.19 \\
        MIKE & 55119.604 & 8.0 & 63.78\,$\pm$\,1.05  \\
        MIKE & 55363.876 & 10.0 & 64.23\,$\pm$\,1.17  \\
        MIKE & 55410.775 & 14.5 & 62.98\,$\pm$\,0.50  \\
        MIKE & 55412.774 & 35.0 & 62.51\,$\pm$\,0.28 \\
        X-shooter & 57961.8453 & 170.0 & 64.60\,$\pm$\,0.55  \\
        X-shooter & 58822.5208 & 120.0 & 66.39\,$\pm$\,0.85  \\
        
        \hline
	\end{tabular}
\end{table}

\section{Abundance Analysis} \label{Abundances}

\begin{figure*}
	\includegraphics[width=1.0\textwidth]{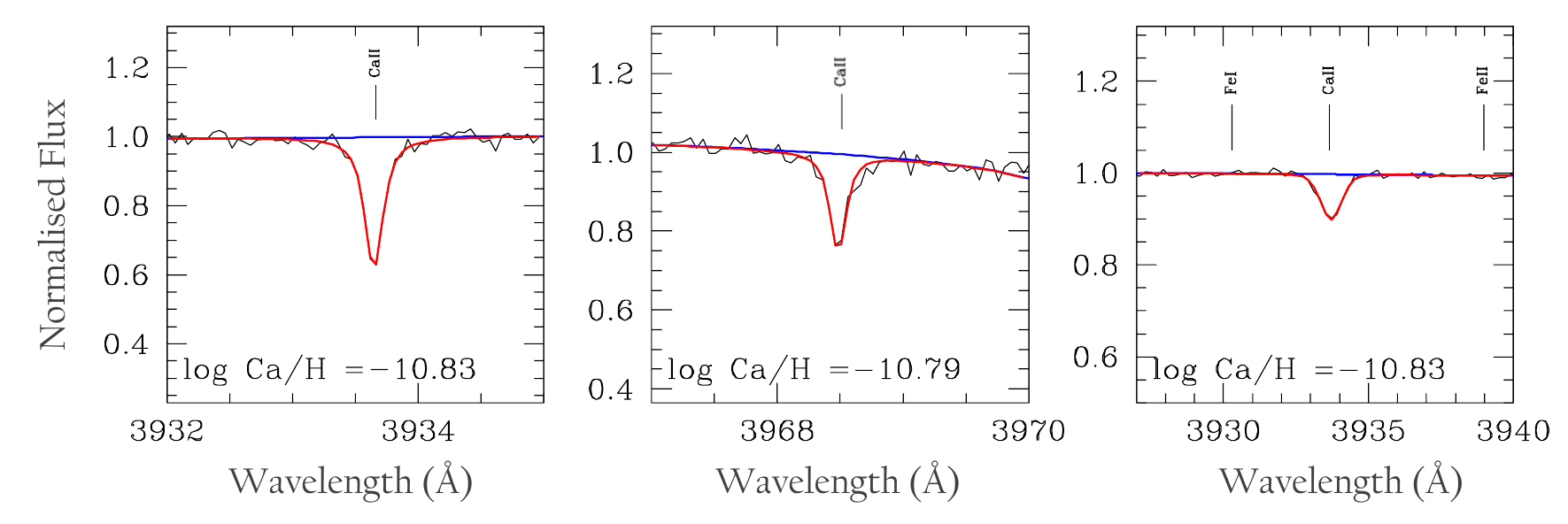}
    \caption{White dwarf model fits (red lines) to the calcium H and K lines from the MIKE data (left and middle), and the calcium K line from the X-shooter data (right). The abundance was calculated per line, with the fitted log(Ca/H) abundance labelled at the bottom of each panel. These values assume the spectroscopic temperature and $\log(g)$ reported in Section \ref{WD-params}.}
    \label{fig:Patrick-fits}
\end{figure*}

\begin{figure}
	\includegraphics[width=\columnwidth]{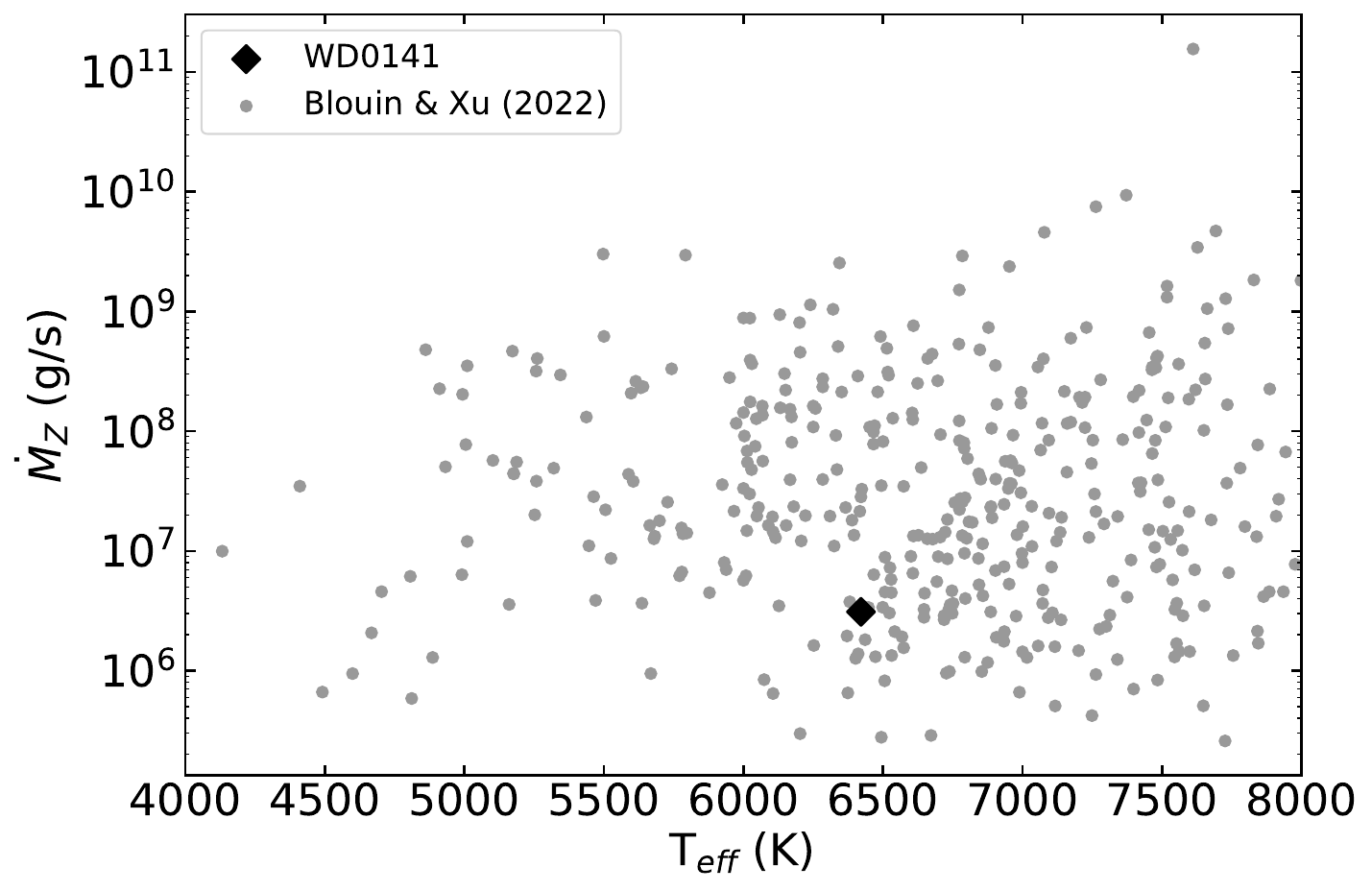}
    \caption{Accretion rate as a function of effective temperature for white dwarfs between 4000 and 8000\,K from \citet{blouin2022no}, calculated assuming the white dwarf is accreting material with bulk Earth-like abundances. WD\,0141$-$675 is over-plotted as a black diamond and sits at the lower end of the distribution.}
    \label{fig:Acc-rate}
\end{figure}

Using the new white dwarf parameters derived in Section \ref{WD-params}, the stacked spectra from the five sets of Magellan MIKE observations, and the VLT X-shooter spectra introduced in Section \ref{Spec_Obs} were fitted separately to obtain the abundance of the polluting planetary material. The [Ca/H] abundance was measured for the stacked MIKE data, and the X-shooter data following methods in \citet{dufour2012detailed}. The spectra were divided into regions of 5--15\,\AA \,\,around the calcium H and K absorption lines. The white dwarf effective temperature and $\log(g)$ were inputted using the spectroscopic and photometric temperature and $\log(g)$ from Table \ref{tab:WD-prop}, and the best-fitting abundance for that spectral line was found; model fits to the data are shown in Fig.\,\ref{fig:Patrick-fits}. Table \ref{tab:Abs} reports the average abundances calculated using the spectroscopic and photometric temperatures. The reported uncertainties contain contributions from the abundance spread from measuring multiple calcium lines including with two instruments, the propagated error in the equivalent width measurements, and error induced from differing white dwarf parameters. Updated diffusion timescales from \citet{koester2014frequency,koester2020new} were used to derive the calcium accretion rate using the spectroscopic and photometric temperature and $\log(g)$, reported in Table \ref{tab:Abs}. The total accretion rate was calculated, assuming that calcium makes up 1.6 percent of the mass of the body as in Bulk Earth \citep{allegre1995chemical}. Figure\,\ref{fig:Acc-rate} shows the accretion rate as a function of effective temperature for white dwarfs between 4000 and 8000\,K from \citet{blouin2022no}. The accretion rate of WD\,0141$-$675 is low compared to the population of polluted white dwarfs of this temperature range. 

\begin{figure}
	\includegraphics[width=1.0\columnwidth]{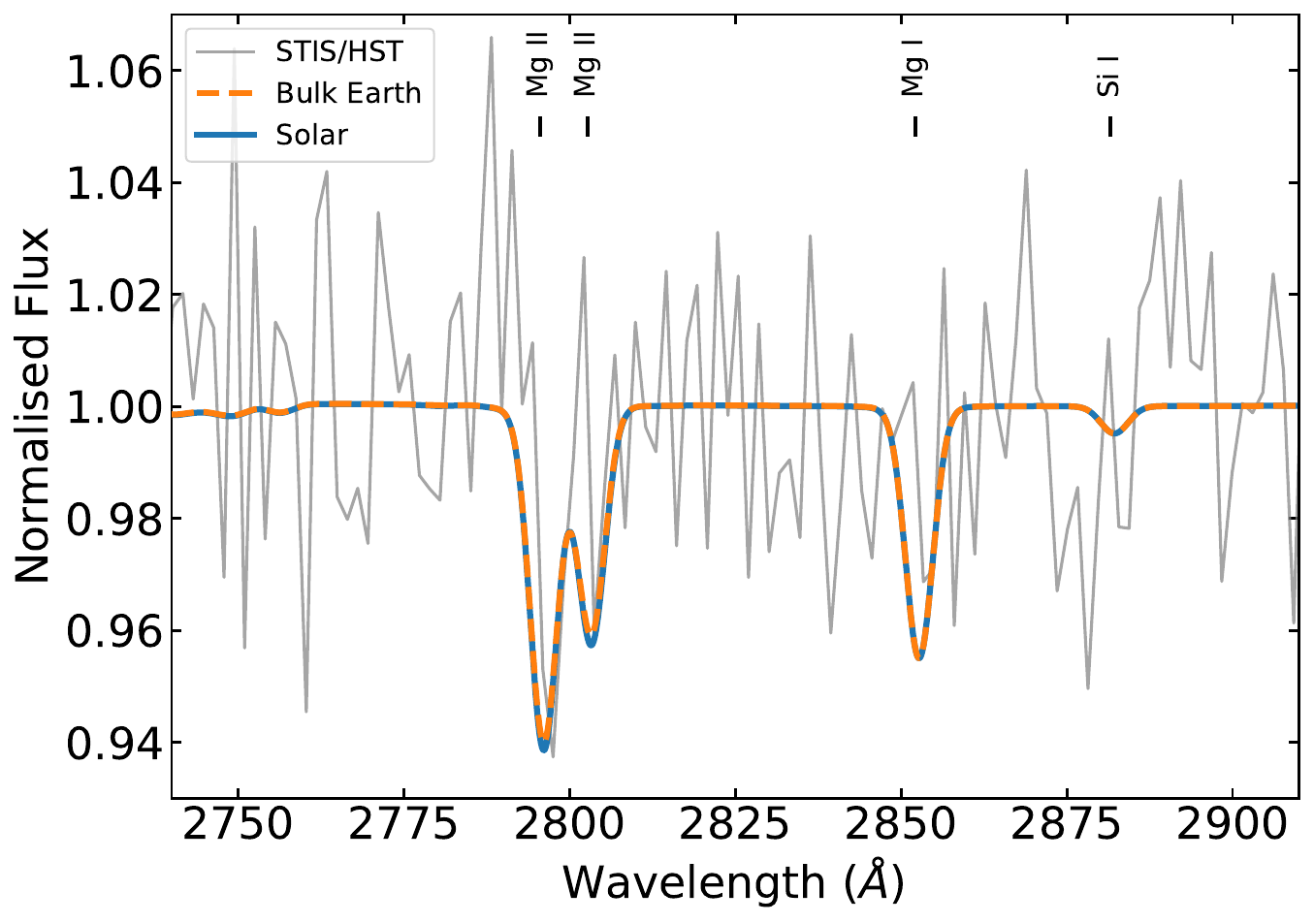}
    \caption{Non detection of Magnesium in the \textit{HST} STIS data. After the calcium lines, assuming either Bulk-Earth or Solar abundances in build-up accretion phase, the next strongest lines in the spectra are Mg I and II lines around 2790--2850\,\AA. No detection of Mg II is made in the STIS data, however, the data are consistent with the models.}
    \label{fig:abundance-fits}
\end{figure}

The spectra are found to be consistent with the accretion of material of either bulk Earth, as may be expected for rocky asteroids, or Solar composition, as may be expected from comets or the outer layers of gas giants. The [Ca/H] abundance was used to scale to Bulk Earth \citep{mcdonough20033} and Solar \citep{grevesse2007solar} abundance, and models were computed for these abundances assuming the spectroscopic temperature and $\log(g)$. The models were compared to the data, and no further features that should have been detected were observed in the data. From the models, the next strongest lines in the spectra are the Mg I and II lines around 2790--2850\,\AA. Figure\,\ref{fig:abundance-fits} shows the predicted models convolved to the resolution of STIS. No detection of Mg II is made in the STIS data, and the spectra are consistent with bulk Earth and Solar-like compositions. Magnesium abundances are similar for bulk Earth and Solar compositions, so to differentiate them the detection of more volatile species such as oxygen or carbon would be required. Additionally, to determine the phase of accretion: build-up, steady-state or declining phase, at least three elements are required covering a range of elemental sinking timescales. Therefore, higher resolution or higher S/N data would be required to confirm whether the pollutant composition was bulk Earth or Solar like, and the phase of accretion.

\begin{table}
	\centering
	\caption{Abundance of the pollutant of WD\,0141$-$675, and inferred accretion rates from calcium, and a total accretion rate assuming a bulk Earth composition (where calcium makes up 1.6 percent of the mass of the body).}
	\label{tab:Abs}
	\begin{tabular}{lll} 
    \hline
     & Spec: & Phot: \\
    \hline
    
    Average [Ca/H]: & $-$10.72$\pm$0.10 & $-$10.82$\pm$0.10 \\
    log($\tau _{\textrm{Ca}}$) (yrs): & 3.94 & 4.24 \\
    $\dot{M} _{\textrm{Ca}}$ (g\,s$^{-1}$): & 5.0$\times 10^4$ & 3.6$\times 10^4$ \\
    $\dot{M} _{\textrm{Total}}$ (g\,s$^{-1}$): & 3.1$\times 10^6$ & 2.2$\times 10^6$ \\
    \hline
	\end{tabular}
\end{table}

\section{Origin of Calcium in the white dwarf} \label{Simulations}

The case study system, WD\,0141$-$675, is a polluted white dwarf, meaning planetary material must have arrived in the photosphere of the white dwarf. This section models WD\,0141$-$675 with a planet on the astrometric orbit and investigates the origin of calcium in the white dwarf, and whether pollution in white dwarfs with close-in astrometric planet candidates is a positive indication that the planet is real.

\subsection{Mass loss from close-in planets} \label{mass-loss}

Atmospheric mass loss, or escape, has been extensively detected in short-period transiting exoplanets \citep[e.g.,][]{Vidal03, Fossati2010, Ehrenreich15, Allart2018, Spake2018, Sing2019}. Most of these are observed in hot gas giants that receive extreme levels of high-energy irradiation, which drives mass loss rates $> 10^9$~g~s$^{-1}$ \citep[e.g.,][]{Erkaev2015, Salz2016}. \citet{gansicke2019accretion} inferred a gas giant planet around WD\,J091405.30+191412.25, this has an expected mass loss rate which matches that of the white dwarf accretion rate, highlighting this scenario as a plausible pollution mechanism in the post main sequence. The energy-limited mass loss formulation \citep[see, e.g.,][]{Salz16a} predicts that the escape rate $\dot{M}$ of a planet with a H/He-dominated atmosphere is given by:

\begin{equation}
    \dot{M} = \frac{\epsilon \pi R_{\rm p}^3 F_{\rm XUV}}{K G M_{\rm p}} \mathrm{,}
\end{equation}where $\epsilon$ is the efficiency of converting irradiation into an outflow, $F_{\rm XUV}$ is the incident XUV flux, $K$ is a correction factor that takes into account the tidal effects \citep{Erkaev07}, $R_{\rm p}$ is planetary radius and $M_{\rm p}$ is the planetary mass. For the test case system of WD~0141$-$675, a planet mass of 9.26~M$_{\textrm{Jup}}$, radius of 1.0~R$_{\textrm{Jup}}$, semi-major axis of $0.172$~au, and an XUV spectral energy distribution as estimated in Section \ref{sect:SED_model}, gives a photoionisation escape rate of just 15~g~s$^{-1}$. In addition, it is unlikely that the planet is losing significant amounts of mass from Jeans escape due to its inferred high gravity and low temperature. Thus, photoionisation and thermal escape from the planet are unlikely to be the cause of the pollution observed in the host star. These models show that for cooler white dwarfs (T$_{\textrm{eff}} < 7,000\,$K), mass loss from close-in giant planets cannot cause significant atmospheric pollution, such as is observable for WD\,0141$-$675, due to the low incident XUV flux. However, for hotter white dwarfs with close-in planets, this effect can explain the accretion rate \citep{gansicke2019accretion}, so must be considered as a potential source of the pollution. 

\subsection{Can scattering supply the observed pollution? } \label{Scattering}

A standard pathway for white dwarf pollution involves planetesimals scattered inwards by planets. The following work considers whether scattering is a viable option to explain the observed accretion. The scenario is considered that the white dwarf is orbited by a close-in giant planet and there is an unseen reservoir of planetesimals in the outer white dwarf planetary system, similar to an asteroid belt that has survived the star's evolution. The material from this outer belt would either be directly scattered by the candidate planetary companion \citep[e.g.][]{ bonsor2011dynamical, debes2012link} or scattered inwards towards the observed planetary companion by further undetected planets \citep[e.g.][]{bonsor2012scattering,marino2018scattering}.

\begin{figure*}
	\includegraphics[width=0.7\textwidth]{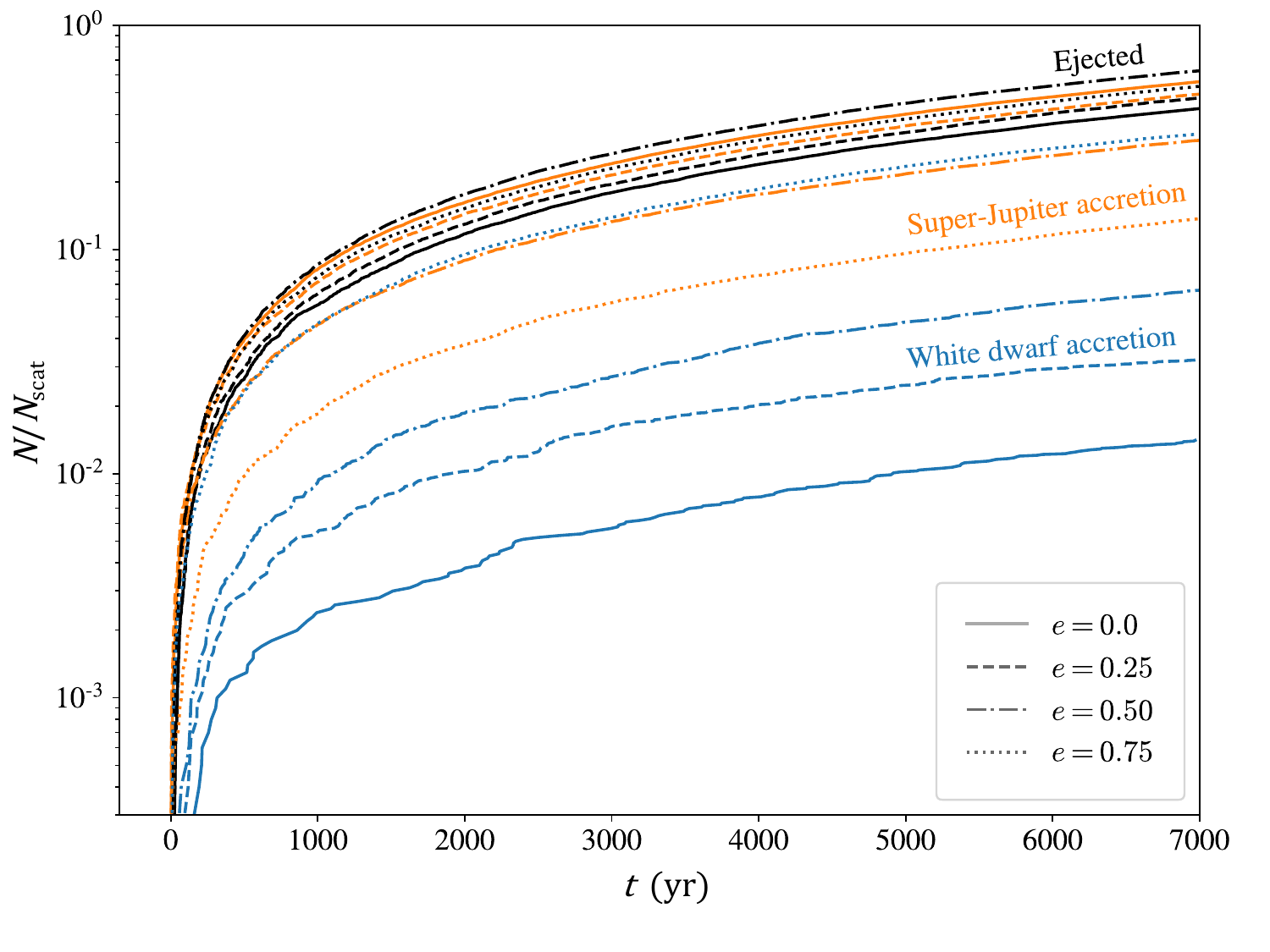}
    \caption{Results from the N body simulations described in Section \ref{Scattering}. The number of particles ejected from the system, accreted onto the super-Jupiter, and accreted onto WD\,0141$-$675 are plotted as a function of time, normalised to the total number of scattered particles. The fractions are plotted for four separate simulations in which test-particles are injected at a constant rate into the chaotic region, $\delta a_\mathrm{chaos}$, of the super-Jupiter with initial eccentricities 0.0, 0.25, 0.50, 0.75 respectively. This behaviour represents the evolution at any time period when particles enter the planet's chaotic zone as the fraction remains unchanged, and Table \ref{tab:simulation-numbers} reports the fractions ejected from the system, accreted onto the super-Jupiter, and accreted onto the white dwarf. }
    \label{fig:simulation}
\end{figure*}

To determine the likelihood of planetesimals being scattered towards the white dwarf and making it past the planet, simulations were conducted using the REBOUND N-body code \citep{rebound}. The MERCURIUS integrator was used \citep{Rein2019}, which uses the high-order adaptive integrator IAS15 during close encounters \citep{rein2015ias15} and otherwise the symplectic integrator WHFAST \citep{WisdomHolman1991,ReinTamayo2015}. In this work mass-less test-particles were injected at a constant rate,  R$_{\textrm{in}}$, into the chaotic region, $\delta a_\mathrm{chaos}$, of the planet, where the chaotic region is defined as $\delta a_\mathrm{chaos} = 1.3 a_\mathrm{pl} \left(\frac{M_\mathrm{pl}}{M_\ast}\right)^{2/7}$ \citep{Wisdom1980Resonance}, and then scattered from this region. The number of particles ejected from the system, and collided with the white dwarf or the planet was tracked, as shown in Fig.\,\ref{fig:simulation}. The criterion for a particle to hit the white dwarf is taken to be when it enters the Roche sphere (approximately 1.0\,R$_{\odot}$), the criterion for a particle to collide with the planet is when it enters the physical radius of the planet (1.0\,R$_{\textrm{Jup}}$), and particles are removed from the simulation when the semi-major axis exceeds 150\,au, thus ensuring that all particles on extreme eccentric orbits can still return and interact with the system. 

Figure\,\ref{fig:simulation} shows that across the wide range of initial particle eccentricities, few particles make it past the planet and hit the white dwarf. Given a constant input rate of particles with a specific orbital eccentricity, the simulations demonstrate that there is an almost constant ratio of particles that are ejected from the system, collide with the Super-Jupiter, and are accreted by the white dwarf, compared to the total number of particles scattered. These scattering rates are determined by the relative probability of each outcome, which is found to be dependent on the eccentricity of the planetesimal belt. The corresponding probability of each scattering outcome are given in Table \ref{tab:simulation-numbers}. The few particles that do make it past the planet may be sufficient to supply the observed accretion. Given the accretion rate of 3.1$\times$10$^6$\,g\,s$^{-1}$ (assuming the spectroscopic white dwarf parameters), over a cooling time of 2.3\,Gyrs, an efficiency of 1 percent (for the $e = $ 0 case), 2.3$\times$10$^{25}$\,g in small bodies must enter the chaotic zone of the inner planet over the cooling age of the white dwarf. This is equivalent to 0.4 percent of Earth's mass or 10 asteroid belts (0.06 Kuiper belts). This assumes that accretion began as the white dwarf formed and continued accreting at the same rate in steady state. However, in reality, this would not be the case, given the stochastic nature of white dwarf planetary systems \citep{wyatt2014stochastic}, it is likely that there were periods of enhanced or no accretion. The most massive observed debris discs contain hundreds of Earth masses \citep{krivov2021solution}, thus, it remains plausible that the observed accretion for this case study was supplied by the scattering of small bodies.

\begin{table}
	\centering
	\caption{Scattering outcomes from the N body simulations described in Section \ref{Scattering}. The table gives the fraction of the different scattering outcomes compared to the total number of particles scattered from the chaotic zone ($\textrm{N}_{\textrm{Scat}}$) across the total time of the simulations. These outcomes are that the particles: (1) get ejected from the system ($\textrm{N}_{\textrm{esc}}$), (2) collided with the Super-Jupiter ($\textrm{N}_{\textrm{SJ}}$), and (3) collided with the white dwarf ($\textrm{N}_{\textrm{WD}}$); these are shown for different eccentricities ($e$) of the disc. Given the accretion rate onto the white dwarf using the spectroscopic and photometric white dwarf parameters (see Table \ref{tab:WD-prop}), the total mass scattered from the disc over the cooling time was calculated (2.3\,Gyrs and 1.95\,Gyrs for the spectroscopic and photometric white dwarf parameters respectively) using M$_{\textrm{$\tau _{\textrm{cool}}$}} = \tau _{\textrm{cool}} \dot{M}_{\textrm{WD}} \frac{\textrm{N}_{\textrm{Scat}}}{\textrm{N}_{\textrm{WD}}}$, where $\dot{M}_{\textrm{WD}}$ is the derived accretion rate onto the white dwarf from Section \ref{Abundances}.}
	\label{tab:simulation-numbers}
	\begin{tabular}{cccccc} 
		\hline
		$e$ & $\frac{\textrm{N}_{\textrm{esc}}}{\textrm{N}_{\textrm{Scat}}}$ & $\frac{\textrm{N}_{\textrm{SJ}}}{\textrm{N}_{\textrm{Scat}}}$ & $\frac{\textrm{N}_{\textrm{WD}}}{\textrm{N}_{\textrm{Scat}}}$ & Spec M$_{\textrm{$\tau _{\textrm{cool}}$}}$\,(g) & Phot M$_{\textrm{$\tau _{\textrm{cool}}$}}$\,(g) \\
        \hline
        0.00 & 0.42 & 0.56 & 0.01 & 2.3$\times$10$^{25}$ & 1.4$\times$10$^{25}$  \\
        0.25 & 0.47 & 0.49 & 0.03 & 7.5$\times$10$^{24}$  & 4.5$\times$10$^{24}$  \\
        0.50 & 0.62 & 0.30 & 0.06 & 3.8$\times$10$^{24}$  & 2.3$\times$10$^{24}$  \\
        0.75 & 0.52 & 0.13 & 0.32 & 7.0$\times$10$^{23}$  & 4.2$\times$10$^{23}$  \\

        \hline
	\end{tabular}
\end{table}

The eccentricity of the disc affects the fraction of bodies that collide with the white dwarf and planet. As the eccentricity increases, the fraction of bodies that collides with the white dwarf increases, whereas the fraction that collides with the planet decreases, as seen in Fig.\,\ref{fig:simulation} and Table \ref{tab:simulation-numbers}. It is only at an eccentricity of 0.75 that the fraction of particles colliding with the white dwarf is greater than the fraction of particles colliding with the planet. These findings agree with previous works on the importance of highly eccentric belts to pollute white dwarfs. From Table \ref{tab:simulation-numbers} it is shown that the planet would also accrete a significant mass of small bodies. Over the cooling age of the white dwarf the planet will accrete 0.01 percent of its mass (9.26\,$M_{\textrm{Jup}}$) in small bodies for the $e=0$ case down to 0.000001 percent of its mass for the $e=0.75$ case. This may result in detectable metallicity enhancements to the atmosphere of the gas giant planet.

Whilst the models reported in Section \ref{mass-loss} indicate that mass loss directly from the atmospheres of close-in giant planets is unlikely to lead to atmospheric pollution for this case study, scattering can continue to supply pollution in systems with close-in planets, given a sufficiently massive reservoir of material. Thus, whilst pollution may not be the natural companion to close-in giant planets \citep[for full details see][]{sanderson2022can}, this work shows that pollution in the atmosphere of a white dwarf with an astrometric planet detection does not suggest that the signal is not real, rather that the white dwarf is orbited by a massive active outer planetary system.

\section{Discussion} \label{Discussion}

Over the lifetime of \textit{Gaia}, it is expected to increase the total number of detected exoplanets by a factor of 14 \citep{perryman2014astrometric}. These astrometric detections provide the full orbital solutions as well as the mass of the exoplanet \citep{holl2022gaia}. However, to get down to planetary masses, high precision astrometry is required and even then, binaries with two similar stellar components can mimic these astrometric signals \citep{2023arXiv230508623M}. Therefore, follow-up observations are crucial to confirm the planetary nature of these astrometric signals. This work presents a case study using the nearby polluted white dwarf WD\,0141$-$675, which had a \textit{Gaia} astrometric planet candidate (now retracted) of mass 9.26\,M$_{\textrm{Jup}}$ on a period of 33.65\,d. The posterior distribution of masses for this companion overlapped into the brown dwarf mass regime, therefore, in order to confirm an astrometric planet candidate it is crucial to obtain follow-up data to place tighter constraints on the mass of the companion. This work collated spectroscopic and photometric observations of WD\,0141$-$675 with an aim of demonstrating how to follow-up astrometric companions to confirm their planetary nature.

\subsection{Confirming the Planetary Nature of a Companion}

One potential false-positive for astrometric planet candidates is a binary, where the two stars are similar - a double degenerate binary with two nearly equal-brightness white dwarfs \citep{collaboration2022gaia,2023arXiv230508623M}. Using the case study of WD\,0141$-$675, this can be ruled out. Firstly, no secondary lines are observed in the high-resolution spectra. Additionally, two independent methods are used to find the best fitting white dwarf parameters, the photometric and spectroscopic methods. If there were two white dwarfs, a larger radii would be needed to explain the luminosity, which would give a much lower $\log(g)$ than a 0.13 dex difference for the photometric method. The radial velocities can provide further constraints on the candidate mass which rules out white dwarf masses on a period of 33.65\,d with an inclination of 87.0 deg. However, caution should be taken as there could be a blend in lines where two stars have overlapping lines and this can create the illusion of no net radial velocity shift \citep{2004ApJBlends}. This can lead to asymmetric line profiles, so to rule this out the shape of the absorption feature should be studied over time. For WD\,0141$-$675 no asymmetric shifts are observed in the absorption line profiles. The re-normalised unit weight error (RUWE) measured from \textit{Gaia} can indicate the presence of close binary companions \citep[e.g.][]{belokurov2020unresolved} by introducing excess scatter into astrometric solutions. Stars with high excess scatter would have RUWE values $>$\,1.25 (if using \textit{Gaia} DR3), and those with average excess scatter would have a RUWE value closer to 1 \citep{penoyre2022astrometric}. For WD\,0141$-$675, the RUWE is 1.049 implying no excess scatter, this means that if there were a companion it only causes a very slight perturbation to the single body fit. The \textit{Gaia} RUWE value is consistent with a single body astrometric solution, implying no large deviation from a massive close-in ($<$\,30\,yr orbit) tertiary companion. As demonstrated by this case study, by collating spectroscopic and photometric data, this false-positive scenario can be ruled out for astrometric planets around white dwarfs. 

\begin{figure*}%
    \centering
    \subfloat[\centering ]{{\includegraphics[trim={0 0cm 0 0},clip,width=0.5\textwidth]{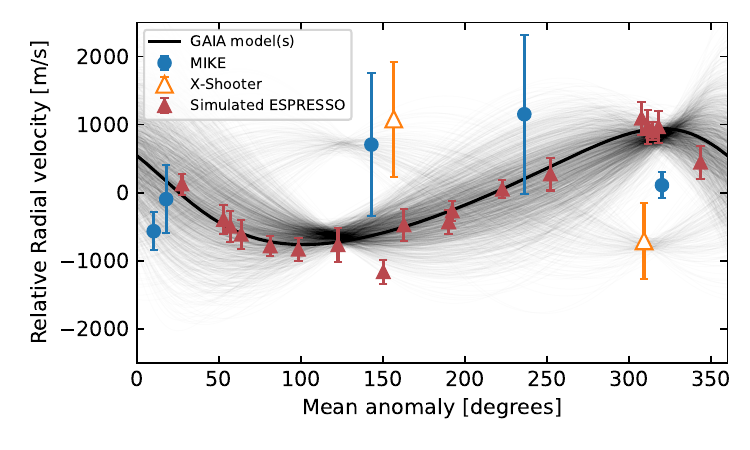} }}%
    \subfloat[\centering  ]{{\includegraphics[trim={0 0.4cm 0 0},clip,width=0.5\textwidth]{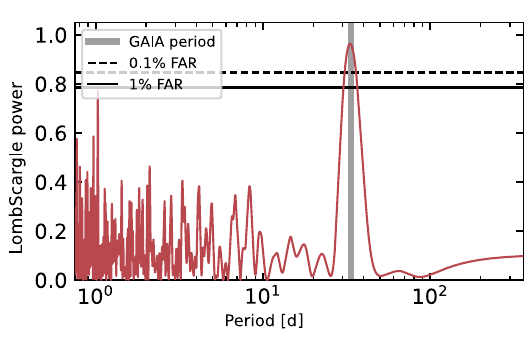} }}%
    \caption{(a) A comparison between the radial velocities presented in this paper, and simulated radial velocities with increased numbers of observations taken at a higher frequency with VLT ESPRESSO. The MIKE and X-Shooter radial velocities phase folded on a 33.65 d period, where the weighted mean values have been subtracted to illustrate the relative radial velocity shift and take any absolute radial velocity shift between the two instruments into account. The black opaque solid line is the \textit{Gaia} best-fit model. The translucent lines are 1000 randomly chosen models from the re-sampled posterior solution of the original astrometric solution. The red triangles are simulated data for a high-resolution stable spectrograph such as VLT ESPRESSO. (b) The Lomb-Scargle periodogram of the 18 simulated ESPRESSO radial velocities. The planet signal is recovered with better than 0.1 percent False Alarm Rate (FAR). }%
    \label{fig:RVs-Louise}%
\end{figure*}

Radial velocities provide a powerful tool to constrain the mass of the object inducing the astrometric signals, if the orbital inclination is favourable. For the case study of WD\,0141$-$675, using sparsely sampled, archival low- and medium-resolution spectra, assuming a nearly edge on orbit from the astrometric signature allows close in (P\,$\sim$\,days) planets $>$\,6\,M$_{\textrm{Jup}}$ to be ruled out (Fig.\,\ref{fig:RVs-model}a). However, with no prior information on the inclination, it is not viable to confidently constrain the mass of a planetary companion (Fig.\,\ref{fig:RVs-model}b). There are significant radial velocity shifts between instruments, as seen from Table\,\ref{tab:RVs}, making the use of archival data with multiple instruments more challenging. Additionally, using instruments that are not designed for precise radial velocity precision means that for this case study, only companions with orbits that produce signals $>$\,1.41\,km\,s$^{-1}$ are detectable. However, with dedicated radial velocity campaigns using a high-resolution stable spectrograph such as VLT ESPRESSO (R$=$140,000 with radial velocity noise floor of 0.5\,m\,s$^{-1}$), constraining companion masses for white dwarf stars is possible \citep{pasquini2023accurate}. The typical ESPRESSO radial velocity uncertainty was assessed by scaling the errors from X-shooter and MIKE with the expected S/N, spectral resolution, and number of useful spectral lines across the wide ESPRESSO wavelength coverage. Here, the increase in spectral resolution is predominantly taken into account as an improved sampling of each spectral line (resulting in a higher S/N). Error-bars of 125$-$250 m\,s$^{-1}$ are achievable with ESPRESSO in 1 hr exposures. The observation campaign simulated covers a baseline of 90\,d (2.5 planetary orbits) where WD\,0141$-$675 is observed with a 5-night cadence, with some randomness added to the sampling to avoid aliases. Figure\,\ref{fig:RVs-Louise} shows 18 simulated radial velocities for ESPRESSO, and the Lomb-Scargle periodogram of this simulated data which clearly shows a detection of a planet with period 33.65\,d. It should be noted, however, that even though the metal lines are narrower than the hydrogen and helium features, broadening effects are still present which give a limit to the radial velocity precision that is independent from the instrument's limit.  

With the successful launch of \textit{JWST}, highly sensitive photometric measurements are possible beyond 10\,\micron\ with the MIRI imager, where the emission of the white dwarf is significantly fainter but still easily detected at high S/N in a reasonable integration time. Using WD\,0141$-$675 as a case study, Fig.\,\ref{fig:Sonora}, especially at F1500W, F1800W, and F2100W, the predicted emission from the companion will be 15, 27, and 47 percent of the white dwarf's emission, while the expected absolute flux calibration accuracy is approximately 2 percent. Therefore, \textit{JWST} is an invaluable tool to help confirm the presence and properties of a planet candidate around a white dwarf. 

\begin{figure}
	\includegraphics[width=1.0\columnwidth]{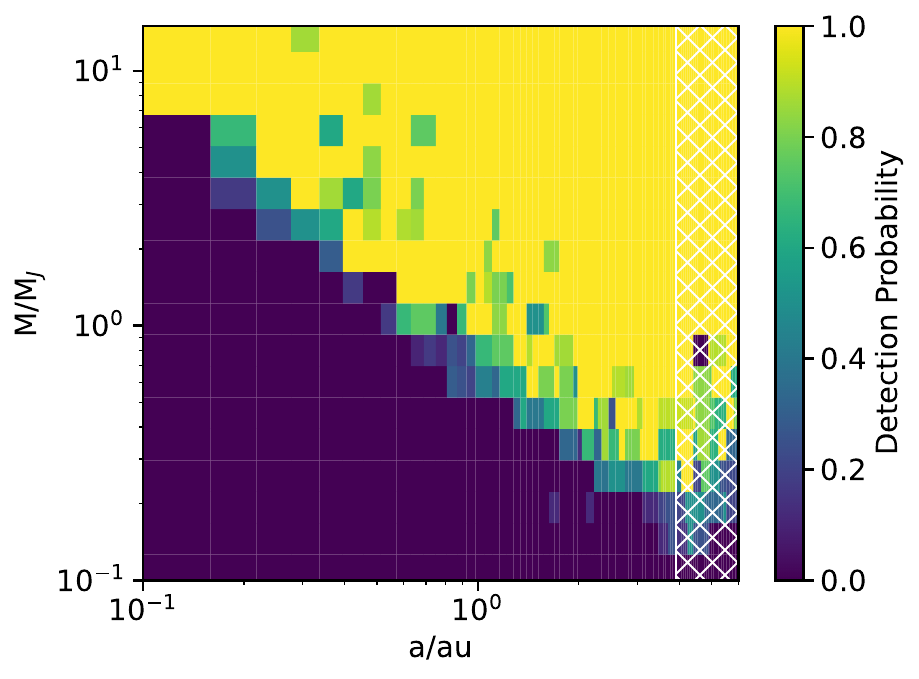}
    \caption{Probability of an astrometric planet detection as a function of mass and semi-major axis for WD\,0141$-$675 using \citet{sanderson2022can}. The white hatched region highlights semi-major axes with periods longer than the \textit{Gaia} mission length. The regions where the detectability drops below one above the yellow boundary are numerical, see \citet{sanderson2022can} for further details.}
    \label{fig:Hannah-Gaia}
\end{figure}

Figure\,\ref{fig:Hannah-Gaia} shows the probability of astrometrically detecting a planet around WD\,0141$-$675 of a given mass and semi-major axes with \textit{Gaia}. If no astrometric signals are observed, planets greater than 7\,$M_{\textrm{Jup}}$ can be ruled out, out to a few au, and outside of 1\,au planets down to 0.5\,$M_{\textrm{Jup}}$ can be ruled out. If an astrometric signal is detected, \textit{JWST} MIRI could detect infrared emission from the companion which can confirm the planetary nature, and targeted radial velocity campaigns can further constrain the mass. For this case study, combining the radial velocity limits with the infrared emission limits, if a cloud-less planet had a 33.65\,d period with an inclination of 87 degrees, the planet would be less than 15.4\,$M_{\textrm{Jup}}$. This is approximately in the planetary mass regime. However, if the planet were further from the white dwarf than predicted, on a less inclined orbit, or has a cloudy atmosphere, this could result in a significant reduction in the strength of the infrared emission of the companion and the amplitude of the radial velocity signature.

\subsection{WD\,0141$-$675 Pollution}

Previous works using the 2017 X-shooter data derived a [Ca/H] abundance of $-$10.96\,$\pm$\,0.11 for WD\,0141$-$675 using a temperature of 6150\,K and $\log(g)$ of 7.58 \citep{kawka2019evidence}. This is consistent within errors to the abundance derived in this work, with the differences likely due to the differing white dwarf parameters used to fit the [Ca/H] abundance.

Based on abundance estimates from the spectra, a total accretion rate of 10$^{6}$ g\,s$^{-1}$ was inferred, the equivalent of a small asteroid's worth of material accreted each year. This is not enough to expect a bright and warm infrared excess due to dust \citep{rafikov2011runaway,rafikov2011metal}, especially given that dust discs around cool white dwarfs are rare, and is consistent with the lack of an infrared excess as seen in Fig.\,\ref{fig:SED}.

\subsection{Is atmospheric pollution expected for white dwarfs with close-in giant planets?}

Spectroscopic observations, such as those described in Section \ref{Spec_Obs}, can indicate whether metals are present in the atmosphere of these white dwarfs. These metals are generally associated with the presence of an outer planetary system. However, in the case of close-in giant planets, it is crucial to consider whether atmospheric pollution is an indicator that supports the presence of the planet.

\cite{sanderson2022can} shows that it is harder to scatter material onto star-grazing orbits in the presence of a close-in giant planet due to the likely dynamical gaps in the planetary system following mass-loss on the giant branch. In Section \ref{Scattering}, this work shows that whilst this is the case, in the scenario that sufficient material is present in the outer planetary system, the relatively low levels of metals observed in the atmosphere of WD\,0141$-$675 could be supplied by scattering, whilst the star was orbited by a planet on the original \textit{Gaia} astrometric solution (see Table \ref{tab:WD-prop} for details). On the other hand, atmospheric pollution from photoionisation atmospheric mass loss and Jeans escape were both found to be implausible to explain the level of accretion onto the white dwarf (see discussion in Section \ref{mass-loss}). Thus, whilst the conclusion remains that on a pollution level, there is unlikely to be as high levels of pollution in white dwarf planetary systems with close-in giant planets compared to those with only lower mass (rocky) planets, it is not possible to use atmospheric pollution to rule out the presence of close in giant planets in an individual white dwarf planetary system.

\section{Conclusions} \label{Conclusions}

\textit{Gaia} is expected to astrometrically reveal of order 10 planets orbiting white dwarfs \citep{perryman2014astrometric,sanderson2022can}. This work shows that follow-up observations, notably photometric and spectroscopic observations are crucial to confirm the planetary nature of such companions. This work focuses on the retracted astrometric planet candidate orbiting WD\,0141$-$675 with planetary mass of 9.26\,M$_{\textrm{Jup}}$ on an orbit with a period of 33.65\,d as a case study. The main conclusions are as follows:

\begin{enumerate}
    \item Infrared observations from \textit{Spitzer} and \textit{WISE} can be used to rule out young, massive planets. \textit{JWST} MIRI is the only infrared instrument with the sensitivity to detect close-in, giant planets of any age orbiting white dwarfs.
    
    \item Archival low- and medium-resolution spectra can be used to obtain radial velocities if the signal is $\gg 1$\,km\,s$^{-1}$. However, short cadence radial velocity time series using high-resolution stable instruments such as VLT ESPRESSO have the power to detect planetary mass objects on close-in orbits around white dwarfs with radial velocity signatures $\ll 1$\,km\,s$^{-1}$. This is however very costly in terms of telescope time, and thus most useful for systems with robust companions.  
    
    \item Double degenerates with nearly equal-brightness companions are a false positive for astrometric planet signals. This is ruled out as an explanation of the astrometric signal for the case study: the lack of spectral features detected from the secondary, radial velocities constrain the binary mass to be much less than a typical white dwarf mass, and fitting the white dwarf parameters with the spectroscopic and photometric method separately reveal consistent parameters. For future detections this kind of analysis will allow the double degenerate false positive to be ruled out.

    \item N body simulations showed that atmospheric pollution from inwards scattering of small asteroids is unlikely, but not impossible in white dwarf planetary systems with close-in giant planets. Therefore, when planets are found around polluted white dwarfs, it should be considered whether it is dynamically and physically feasible to pollute the white dwarf with small bodies.  

    \item WD\,0141$-$675 is being polluted at a rate of 10$^6$ g\,s$^{-1}$ the equivalent of a small asteroid's worth of material accreted each year. Scaling the calcium abundance ([Ca/H] = $-$10.72) to bulk-Earth and Solar abundances, no additional features should have been detected, meaning the spectra are consistent with both bulk-Earth and Solar. Additional high-S/N and high-resolution data would be required to determine further details on the composition of the pollutant. 

\end{enumerate}

Given that so few planet candidates are known around white dwarfs, every confirmation gives crucial insights into these evolving planetary systems and the future of planetary systems. In particular, which planets survive post-main sequence evolution and which ones do not; this provides information on the efficiency of common envelope evolution for low mass companions, impacting both occurrence rates of sub-stellar objects as well as binary evolution at high mass ratios. Therefore, following up astrometric planet candidates identified with \textit{Gaia} around white dwarfs is crucial.

\section*{Acknowledgements}

We thank the referee for helpful comments that improved the manuscript. LR acknowledges support of an ESA Co-Sponsored Research Agreement No. 4000138341\textbackslash 22\textbackslash NL\textbackslash GLC\textbackslash my = Tracing the Geology of Exoplanets. AB and LR acknowledge support of a Royal Society University Research Fellowship, URF\textbackslash R1\textbackslash 211421.  RJA is grateful for a PhD Studentship from the Science and Technology Facilities Council (STFC). SLC acknowledges support from an STFC Ernest Rutherford Fellowship: ST\textbackslash R003726\textbackslash 1. ZP acknowledges that this project has received funding from the European Research Council (ERC) under the European Union's Horizon 2020 research and innovation programme (Grant agreement No. 101002511 - VEGA P). SX is supported by the international Gemini Observatory, a program of NSF's NOIRLab, which is managed by the Association of Universities for Research in Astronomy (AURA) under a cooperative agreement with the National Science Foundation, on behalf of the Gemini partnership of Argentina, Brazil, Canada, Chile, the Republic of Korea, and the United States of America. This project has received funding from the European Research Council (ERC) under the European Union's Horizon 2020 research and innovation programme (Grant agreement No. 101020057). MK acknowledges support by the National Science Foundation under grant AST-2205736 and NASA under grant 80NSSC22K0479.

This work has made use of data from the European Space Agency (ESA) mission {\it Gaia} (\url{https://www.cosmos.esa.int/gaia}), processed by the {\it Gaia} Data Processing and Analysis Consortium (DPAC, \url{https://www.cosmos.esa.int/web/gaia/dpac/consortium}). Funding for the DPAC has been provided by national institutions, in particular the institutions participating in the {\it Gaia} Multilateral Agreement. This research made use of pystrometry, an open source Python package for astrometry timeseries analysis \citep{2019zndo3515526S}.

\section*{Data Availability}

VLT X-shooter data available from the ESO archive (\href{http://archive.eso.org/eso/eso\_archive\_main.html}{http://archive.eso.org/eso/eso\_archive\_main.html}). Magellan MIKE spectra are available on reasonable request. The TESS data are available from the Mikulski Archive for Space Telescopes (\href{https://archive.stsci.edu/tess/}{https://archive.stsci.edu/tess/}). \textit{Spitzer} data available from the NASA/IPAC Infrared Science Archive (\href{https://irsa.ipac.caltech.edu/Missions/spitzer.html}{https://irsa.ipac.caltech.edu/Missions/spitzer.html}). \textit{HST} STIS data are available from the Mikulski Archive for Space Telescopes  (\href{https://mast.stsci.edu/search/ui/\#/hst}{https://mast.stsci.edu/search/ui/\#/hst}).




\bibliographystyle{mnras}
\bibliography{Refs} 






\bsp	
\label{lastpage}
\end{document}